\PassOptionsToPackage{table,xcdraw}{xcolor}

\documentclass[manuscript, acmlarge,screen]{acmart}

\usepackage{balance}       
\usepackage{graphics}      
\usepackage{hyperref}
\usepackage{color}
\usepackage{longtable}
\usepackage{booktabs}
\usepackage{textcomp}
\usepackage{subcaption}
\usepackage{enumerate}
\usepackage{xcolor}
\usepackage{lipsum}
\usepackage{makecell}
\usepackage{multicol}
\usepackage{multirow}
\usepackage{array}
\usepackage{verbatimbox}
\usepackage{enumitem}
\usepackage{amsmath}
\usepackage{stfloats}
\usepackage{graphicx}
\usepackage{amsthm}
\usepackage{listings}
\usepackage{caption} 
\usepackage[export]{adjustbox}
\usepackage{xspace}
\usepackage{epsfig}
\usepackage[linesnumbered]{algorithm2e}
\usepackage{algpseudocode}
\usepackage{tabularx}
\usepackage{arydshln}
\usepackage[bottom]{footmisc}
\usepackage{tcolorbox}
\usepackage{stackengine}
\usepackage{lipsum}
\usepackage{enumitem}
\usepackage{colortbl}
\usepackage{enumitem}
\setlist{
  listparindent=\parindent,
  parsep=0pt,
}
\usepackage{tikz}
\usepackage{pifont}
\newcommand{\cmark}{\ding{51}}%
\newcommand{\xmark}{\ding{55}}%

\newcommand*{\eg}{\textit{e.g.},\xspace}
\newcommand*{\ie}{\textit{i.e.},\xspace}

\newcommand*{\etal}{\textit{et~al.}\xspace}

\newcolumntype{L}[1]{>{\raggedright\let\newline\\\arraybackslash\hspace{0pt}}m{#1}}
\newcolumntype{C}[1]{>{\centering\let\newline\\\arraybackslash\hspace{0pt}}m{#1}}
\newcolumntype{R}[1]{>{\raggedleft\let\newline\\\arraybackslash\hspace{0pt}}m{#1}}

\algnewcommand{\IfThenElse}[3]{
  \State \algorithmicif\ #1\ \algorithmicthen\ #2\ \algorithmicelse\ #3}

\definecolor{darkblue}{HTML}{0C0893} 
\definecolor{brilliantlavender}{rgb}{0.6, 0.4, 0.8}
\definecolor{candypink}{rgb}{0.89, 0.44, 0.48}
\definecolor{lightpurple}{rgb}{0.8, 0.5, 0.98}
\definecolor{green}{rgb}{0.08, 0.47, 0.16}
\definecolor{violet}{rgb}{0.96, 0.5, 0.5}
\definecolor{asparagus}{rgb}{0.53, 0.66, 0.42}
\definecolor{darkpastelpurple}{rgb}{0.59, 0.44, 0.84}
\definecolor{mediumslateblue}{rgb}{0.48, 0.41, 0.93}

\definecolor{DarkGreen}{HTML}{5DAC81}
\begin{document}

%

\title{Illuminating the Unseen: Investigating the Context-induced Harms in Behavioral Sensing}

%

\author{Han Zhang}
\email{micohan@cs.washington.edu}
\affiliation{%
  \institution{University of Washington}
  \country{USA}
 }

 \author{Vedant Das Swain}
\email{v.dasswain@northeastern.edu}
\affiliation{%
  \institution{Northeastern University}
  \country{USA}
 }

\author{Leijie Wang}
\email{leijiew@cs.washington.edu }
\affiliation{%
  \institution{University of Washington}
  \country{USA}
 }

\author{Nan Gao}
\email{nangao@tsinghua.edu.cn}
\affiliation{%
  \institution{Tsinghua University}
  \country{China}
 } 

\author{Yilun Sheng}
\email{ylsheng@cs.washington.edu}
\affiliation{%
  \institution{University of Washington}
  \country{USA}
 }

  \author{Xuhai Xu}
\email{xoxu@mit.edu}
\affiliation{%
  \institution{Massachusetts Institute of Technology}
  \country{USA}
 }

  \author{Flora D. Salim}
\email{flora.salim@unsw.edu.au}
\affiliation{%
  \institution{University of New South Wales}
  \country{Austrilia}
 }

 \author{Koustuv Saha}
\email{ksaha2@illinois.edu}
\affiliation{%
  \institution{University of Illinois Urbana-Champaign}
  \country{USA}
 } 

\author{Anind K. Dey}
\email{anind@uw.edu}

 \author{Jennifer Mankoff}
\email{jmankoff@cs.washington.edu}
\affiliation{%
  \institution{University of Washington}
  \country{USA}
 }

%

%
\begin{abstract}
Behavioral sensing technologies are rapidly evolving across a range of well-being applications. Despite its potential, concerns about the responsible use of such technology are escalating. In response, recent research within the sensing technology has started to address these issues. While promising, they primarily focus on broad demographic categories and overlook more nuanced, context-specific identities. These approaches lack grounding within domain-specific harms that arise from deploying sensing technology in diverse social, environmental, and technological settings. Additionally, existing frameworks for evaluating harms are designed for a generic ML life cycle, and fail to adapt to the dynamic and longitudinal considerations for behavioral sensing technology. To address these gaps, we introduce a framework specifically designed for evaluating behavioral sensing technologies. This framework emphasizes a comprehensive understanding of context, particularly the situated identities of users and the deployment settings of the sensing technology. It also highlights the necessity for iterative harm mitigation and continuous maintenance to adapt to the evolving nature of technology and its use. We demonstrate the feasibility and generalizability of our framework through post-hoc evaluations on two real-world behavioral sensing studies conducted in different international contexts, involving varied population demographics and machine learning tasks. Our evaluations provide empirical evidence of both situated identity-based harm and more domain-specific harms, and discuss the trade-offs introduced by implementing bias mitigation techniques. 

\end{abstract}

%
%


%

\renewcommand{\shorttitle}{Investigating the Context-induced Harms in Behavioral Sensing}
\renewcommand{\shortauthors}{Zhang et al.}

%
\maketitle

\section{Introduction}
\label{sec:introduction}

The rapid evolution of sensing technologies has unlocked new possibilities for tracking and understanding human activities. Behavioral sensing technology, which involves using sensing to capture, model, and predict human behaviors, offers a broad spectrum of applications. These include, but are not limited to, well-being monitoring (\eg mental health prediction~\cite{morshed2019prediction,adler2020predicting,das2022semantic,saha2021person}), human activity recognition (\eg identifying activities like ``running'', ``sitting'', and ``walking''~\cite{kwon2020imutube, peng2018aroma}), and personalized recommendations (\eg personalized music and taxi charging recommendation systems ~\cite{li2022towards,wang2020faircharge}). This technology, in contrast to the traditional manual approach of using questionnaire-collected data for the same tasks, facilitates continuous, automated, and unobtrusive gathering of \textit{context}~\cite{das2023algorithmic,cornet2018systematic}. Here, context refers to capturing all information related to the interactions among users, applications, and their environment~\cite{dey2001understanding,dey2001conceptual}. 

In recent years, concerns about the responsible use of behavioral sensing technology have been growing~\cite{chowdhary2023can,kawakami2023sensing,kaur2022didn,corvite2023data}. These concerns arise from the prevalent top-down design approach, which typically involves technology builders---a collective of researchers, designers, developers, and engineers---developing tools based on their assumptions of users' goals, needs, or preferences~\cite{mohr2017three,kawakami2023sensing}. This approach often leads to a lack of sufficient understanding of the users' diverse backgrounds and the situation in which the technology is used and deployed, a concept described as \textit{context sensitivity}~\cite{davies1998developing,dey2001understanding}. Consequently, important contextual factors that may not initially appear relevant to its primary purpose are often disregarded during the early design phases. Such neglect can lead to technologies that fail to adequately meet the diverse and real-world needs of users effectively and may even introduce potential harms~\cite{yfantidou2023beyond}. 

Understanding the differences in model performance and the potential harms of sensing technology is critical, as also highlighted in prior research~\cite{yfantidou2023beyond,meegahapola2023generalization,assi2023complex,zhang2023framework,adler2024measuring,yfantidou2023uncovering}. Researchers have started exploring how models may perform differently based on various personal sensitive attributes, a contextual factor we refer to as \textit{identity-based harm}. These attributes include demographics, socioeconomic status, country of residence, and health conditions~\cite{yfantidou2023uncovering,adler2024measuring,meegahapola2023generalization}. Accordingly, we recognize three gaps in the existing literature. 

\textbf{Gap 1}: 
There is a limited understanding of more \textit{situated identities} relevant to specific contexts, such as first-generation college students, immigration status, or disability status---groups that also confront profound societal inequalities~\cite{wilbur2016first,zhang2022impact,behtoui2010social}. The under-representation of these identities fails to capture a more comprehensive spectrum of user experiences and needs within sensing technology design and application.

\textbf{Gap 2}: 
A distinct aspect of sensing technology lies in what we term as \textit{situation-based harm}, which remains understudied. This type of harm emerges specifically from the inherent versatility of sensing technology, which finds applications across a diverse array of scenarios—from varied technological infrastructures and environmental conditions to different device types~\cite{abowd2012next,yau2004adaptive}. This diversity in application settings can lead to unique contextual harms when the technology is designed or evaluated without thorough consideration of these situational differences. Such harms are particularly concerning because they often do not intuitively link directly to users and can go unanticipated, thereby increasing the risk of irresponsible technology deployment and exacerbating the potential for ultimately negative impacts on users. 

\textbf{Gap 3}: 
Understanding the intricacies of bias mitigation, particularly the potential trade-offs introduced by adopting such techniques, remains underexplored. Moreover, the current evaluation frameworks do not adequately cater to the unique demands of sensing technology pipelines. These pipelines require iterative bias mitigation and continuous maintenance to adapt to the dynamic nature of sensing data, challenges that are not fully addressed by existing frameworks (\eg~\cite{suresh2019framework,hutiri2022bias,liu2023reimagining}). Moreover, there is a lack of a systematic approach to effectively design, build, and evaluate sensing technologies, leaving gaps in how these sensing technologies are developed and how their impacts are assessed across different user groups and usage scenarios.

To address the identified gaps in existing research, we propose a theory-driven, human-centered framework specifically designed for evaluating sensing technology. This framework builds upon evaluation frameworks in machine learning~\cite{hutiri2022bias,suresh2019framework,liu2023reimagining} by incorporating a thorough assessment of both broad and domain-specific context-induced harms to users during the evaluation process. Moreover, our proposed framework emphasizes the need for iterative harm mitigation and continuous maintenance, critical components that are essential to accommodate the dynamic nature of sensing technology. 

We conduct post-hoc evaluations on two real-world behavioral sensing studies to demonstrate the feasibility of our proposed framework and to derive new insights from the evaluation results. To ensure the generalizability of our framework, we select two studies conducted in different countries, with different populations, and involving distinct ML tasks. Specifically, one study focused on college student mental health detection~\cite{xu2022globem}, a classification task; and another on teenage student learning engagement prediction~\cite{gao2020n}, a regression task. 

The first evaluation provides empirical evidence of situated identity-based harm, extending our understanding beyond demographic categories such as race and age. The second evaluation, showcases robustness to identity-based harm, but instead, highlights evidence for situation-based harm by showing  significant performance differences in algorithms across environmental situations, such as varying temperature conditions.  Furthermore, our experiment on a bias mitigation technique reveals a critical trade-off: efforts to mitigate harm for one attribute inadvertently introduced harm to other attributes. In summary, our contributions are as follows.

\begin{itemize}
    \item In Section~\ref{sec:framework}, we propose a framework that integrates consideration for both identity-based and situation-based harms, and emphasizes the need for iterative bias mitigation and continuous maintenance to effectively manage the dynamic challenges posed by sensing technologies.
    \item In Sections~\ref{sec:case_study1} and \ref{sec:case_study2}, We conduct evaluations on two real-world behavioral sensing studies. We provide empirical evidence on the nuanced forms of identity-based harms and highlight situational dependencies that affect algorithm performance. Furthermore, we make our analysis codebase openly accessible for reproducibility\footnote{We will release our codebase at publication.}. 
    \item In Section~\ref{sec:discussion}, we offer key insights from our evaluation studies, focusing on the potential harms linked to more situated identity-based factors and the newly observed situation-based harms in behavioral sensing. Additionally, we discuss the trade-offs introduced by implementing bias mitigation techniques.
\end{itemize}

We further offer reflections within and beyond our proposed framework. Our work aims to contribute both conceptually and practically to the field, focusing on more responsible behavioral sensing technologies. 

\section{Background and Related Work}
\label{sec:background}

As behavioral sensing technologies increasingly become a tool for tracking and reasoning about human activities, they present a blend of promising opportunities and potential risks~\cite{zhang2023framework}. In this section, we first review behavioral sensing technology, exploring its evolution and current landscape (Section~\ref{sec:related_work-sensing_tech}). We then review a promising application domain of these technologies, specifically in well-being prediction (Section~\ref{sec:related_work-wellbeing_prediction}). Following this, we discuss the potential harms associated with these technologies, including both identity-based and situation-based harms, particularly arising from a lack of context-sensitivity (Section~\ref{sec:related_work-potential_harms}). We conclude this section by reviewing the human-centered design approach, aimed at addressing these harms (Section~\ref{sec:related_work-existing-approaches}). This background forms the basis for proposing our evaluation framework.

\subsection{Evolution of Behavioral Sensing Technology}\label{sec:related_work-sensing_tech}
Since the late 1990s, researchers in sensing technology have increasingly recognized the importance of enabling computing devices to enhance application performance by incorporating knowledge of the context in which they are used~\cite{abowd1999towards,dey2001understanding,schilit1994context}. ``Context'' in this sense refers to all information related to the interactions among users, applications, and their environment~\cite{dey2001conceptual}. Alongside this realization, there was growing advocacy for the creation of sensing systems designed to offer information or services that are relevant to the specific tasks of users, a concept known as \textit{context-awareness}~\cite{abowd1999towards,dey2001understanding}. Building upon this foundational concept, research efforts have since concentrated on creating various toolkits and frameworks to capture, infer, and generate context through diverse sensors~\cite{ferreira2015aware,lim2010toolkit,dey2001conceptual,salber1999context}. These initiatives have evolved to focus on employing these toolkits for passive data collection, aiming to infer human behavior, such as their phone usage, location, sleep, and steps~\cite{doryab2018extraction}. More recently, the integration of Machine Learning (ML) and Artificial Intelligence (AI) algorithms into these technologies has further transformed this field. Researchers have begun integrating these advanced techniques into sensing technologies, not only for modeling human behavior but also for making predictions~\cite{banovic2016modeling,adler2020predicting,das2020modeling}. 

\subsection{Well-being Predition in Behavioral Sensing}\label{sec:related_work-wellbeing_prediction}

Well-being prediction is one of the promising and extensively studied application domains for behavioral sensing technologies. This domain includes various aspects such as predicting mental health~\cite{adler2020predicting, chikersal2021detecting, sefidgar2019passively} and forecasting performance, engagement, as well as productivity~\cite{wang2015smartgpa, gao2020n}.  Specifically, in the area of mental health prediction, considerable research has been dedicated to depression prediction. For instance, studies have utilized passively sensed data such as physical activities, phone usage, sleep patterns, and step counts of participants to predict depressive symptoms~\cite{wang2018tracking, xu2019leveraging, chikersal2021detecting}. Additionally, there have been significant research efforts aimed at understanding mood-related health concerns among students and workers, employing various types of passively sensed data~\cite{saha2017inferring,morshed2019prediction, meegahapola2023generalization, li2020extraction}. 

In parallel, evaluating performance, engagement, and productivity as a facet of mental health-related well-being prediction has also attracted considerable attention. Many studies have focused on student populations. For example,~\citeauthor{wang2015smartgpa} conducted a study using passively collected data from smartphones and wearables of college students to predict their cumulative GPA. In another work,~\citeauthor{ahuja2019edusense} developed a classroom sensing system to capture student facial expressions and body gestures from audio and video data. Their approach allowed for the analysis of students' engagement levels based on these sensory inputs. In a different study, Gao \etal~\cite{gao2020n,gao2022understanding} employed indoor environmental data, such as temperature, humidity, CO2 levels, and sounds, alongside physical activity data, to predict three dimensions of student engagement levels. Beyond the academic setting, behavioral sensing technology is increasingly being used in the corporate sector to assess workplace productivity and employee well-being~\cite{rahaman2020ambient}. For example, Mirjafar~\cite{mirjafari2019differentiating} trained machine learning algorithms on sensing data to differentiate performance levels in workplaces, offering insights for workspace optimization and stress management~\cite{das2020modeling}.


\subsection{Potential Harms in Behavioral Sensing}\label{sec:related_work-potential_harms} While these behavioral sensing applications offer significant opportunities to improve and support human well-being, they often employ a top-down design approach, which is predominantly driven by technology builders' assumptions of users' goals, needs, or preferences~\cite{mohr2017three,brodie2018big}. However, developing technologies based solely on these assumptions, alongside what is easily possible to sense, without a thorough understanding of the users' diverse backgrounds and the situation in which the technology is used and deployed, can lead these technologies to a lack of \textit{context sensitivity}. In this paper, we expand context sensitivity, traditionally linked to context-awareness~\cite{davies1998developing,dey2001understanding}, and redefine it to highlight the responsible aspects of these technologies regarding diverse user groups and situations.

\begin{quotation}
 \textit{``A technology is \textbf{context-sensitive} when it  accounts for diverse user backgrounds, needs, and situations of use to provide value to users.''}     
\end{quotation}

As behavioral sensing technologies advance toward practical, real-world applications, it becomes increasingly important to ensure that these technologies are context-sensitive. This consideration is crucial to mitigate potential harms to users. Below, we review existing work on one type of harms, identity-based harm, in sensing technology and introduce a domain-specific harm, situation-based harm.

\subsubsection{Identity-based Harms} As highlighted by~\citeauthor{yfantidou2023beyond} in their review of sensing technology research from 2018 to 2022, a mere 5\% of studies investigated algorithmic harms to users with sensitive attributes. Alarmingly, 90\% of these studies limited their focus to only gender and age, and primarily relied on accuracy or error metrics for evaluation. In recent years, Ubicomp researchers have begun to understand the potential harms of sensing technology~\cite{yfantidou2023beyond,meegahapola2023generalization,assi2023complex,zhang2023framework,adler2024measuring,yfantidou2023uncovering}. For example, evaluations conducted on three existing datasets, based on an established framework to identify and understand sources of identity-based harm throughout the ML life cycle~\cite{suresh2021framework}, have provided empirical evidence of identity-based harms across data generation, model building, and implementation processes~\cite{yfantidou2023uncovering}.~\citeauthor{adler2024measuring} showed that sensed-behaviors that predict depression risk were inconsistent across demographic and socioeconomic subgroups. In addition, researchers investigated into how models perform across different countries or cultures—factors that are also tied to identity—have shed light on variations in applications such as activity recognition~\cite{assi2023complex} and mood inference~\cite{meegahapola2023generalization}.  While these developments are promising, these works primarily focused on broad demographic categories and overlooks more nuanced, context-specific identities. This oversight is concerning, because people with these attributes have a history of facing societal inequalities, as extensively documented in psychology and social work research~\cite{schmitt2002meaning, frost2011social,karlsen2002relation,chou2012perception}. For example,~\citeauthor{hangartner2021monitoring} found in their study of online recruitment platforms that individuals from immigrant and minority ethnic groups received 4-19\% fewer recruiter contacts. Similarly,~\citeauthor{blaser2019perspectives} noted the significant absence of disability reporting in tech companies and their media coverage. Moreover,~\citeauthor{erete2021can} employed autoethnography~\cite{ellis2011autoethnography} and testimonial authority~\cite{collins2019intersectionality} to share their experiences as Black women academics during a pandemic disproportionately impacting their communities and in the context of civil unrest due to racial injustice.

\subsubsection{Situation-based Harm}\label{sec:situation-bias} 
Another potential harm that could emerge due to a lack of context sensitivity in behavioral sensing technology is what we identify as situation-based harm. This type of harm could occur when sensing technology is implemented in diverse situations or settings. As this aspect of potential harm is relatively under-explored, we provide an example to help readers conceptually understand it. Specifically, if a behavioral sensing algorithm is predominantly based on data from iOS-based smartphones, it may not be effective on Android-based smartphones due to representational bias~\cite{mehrabi2021survey}. This may potentially lead to a disproportionate impact on individuals of lower socioeconomic status or those in developing countries who commonly use more affordable Android devices. Reportedly, iOS-based smartphones tend to be more than twice as expensive as their Android-based counterparts~\cite{ndibwile2018smart4gap, jamalova2019comparative}.

\subsection{Human-Centered Approach for Investigating Harms}\label{sec:related_work-existing-approaches}

The broader ML community has developed various frameworks to evaluate bias~\cite{suresh2019framework,hutiri2022bias,liu2023reimagining,raji2020closing}. However, these frameworks often fall short in prioritizing users' needs. To identify and mitigate potential harms to humans, there is an increasing call within the HCI, CSCW, and Ubicomp communities for a focus on \textit{human-centered AI} (HCAI) design~\cite{amershi2019guidelines,lee2020human,tahaei2023human,wang2023designing,ethicsEC,acemoglu2021harms,ehsan2023charting}. While definitions of HCAI vary~\cite{capel2023human}, the central theme revolves around designing AI technologies that are deeply attuned to the needs, values, and agency of human users, partners, and operators~\cite{capel2023human}. This approach focuses on ensuring that AI systems are technically efficient and align with human-centric values and ethical standards. Research efforts have been made to provide guidelines for designers to create more ethical AI systems. For example,~\citeauthor{amershi2019guidelines} distilled over 150 AI-related design recommendations into 18 broadly applicable guidelines, including ensuring AI systems are cautious about social biases and enabling users to provide feedback during regular interaction with the AI systems. Adopting this approach, researchers have actively engaged with stakeholders to gain an in-depth understanding of their experiences and perceptions regarding AI in general and sensing technology in particular. This engagement has focused on various aspects, including stakeholders' trust in AI systems~\cite{ma2023should, kim2023humans, vodrahalli2022humans, banovic2023being}, their privacy concerns related to the use of sensing technologies~\cite{reynolds2004affective, rooksby2019student, adler2022burnout}, and the specific impacts of these technologies in different settings~\cite{kawakami2023sensing, chowdhary2023can, corvite2023data, das2023algorithmic, das2024sensible}.

\textit{Value-sensitive} design is another widely adopted approach that can address potential harms to humans~\cite{friedman1996value}. This approach is grounded in the principle of integrating human values into the design process thoroughly and systematically. It utilizes a tripartite methodology that is both integrative and iterative, involving conceptual, empirical, and technical investigations~\cite{friedman2013value,friedman2017survey}. To explicate this approach,~\citeauthor{friedman2013value} presented three case studies in their work. Leveraging this concept, \citeauthor{zhu2018value} applied value-sensitive design to algorithm development. In their approach, they actively engaged with stakeholders during the early stages of algorithm creation, incorporating their values, knowledge, and insights to ensure that the technology aligns with user needs and ethical standards.


\section{An Evaluation Framework for Investigating Context-induced Harms in Behavioral Sensing}
\label{sec:framework}


\subsection{Our Evaluation Perspective: Human-Centered and Context-Sensitive AI}\label{sec:framework-design}

Our evaluation approach is grounded in the principles of \textit{human-centered AI} (HCAI)~\cite{shneiderman2020human,jordan2019artificial} and \textit{value-sensitive} design~\cite{friedman2013value,zhu2018value}, which advocates for a shift away from traditional practices where behavioral sensing technology builders rely on predetermined assumptions about user needs and preferences~\cite{mohr2017three}. Rather, it emphasizes a thorough consideration of potential harms to users throughout the technology design process. This aims to ensure that the behavior of the developed technologies does not reinforce negative stereotypes or biases against users, in line with the guidelines proposed by Amershi \etal\cite{amershi2019guidelines}. Moreover, we emphasize the importance of iteratively mitigating potential harms to users and maintaining continuous oversight to address the dynamic nature of sensing technology~\cite{zhang2023framework}.

\subsection{Overview of the Framework}\label{sec:framework-overview}

\begin{figure}
    \centering
    \includegraphics[width=1\textwidth]{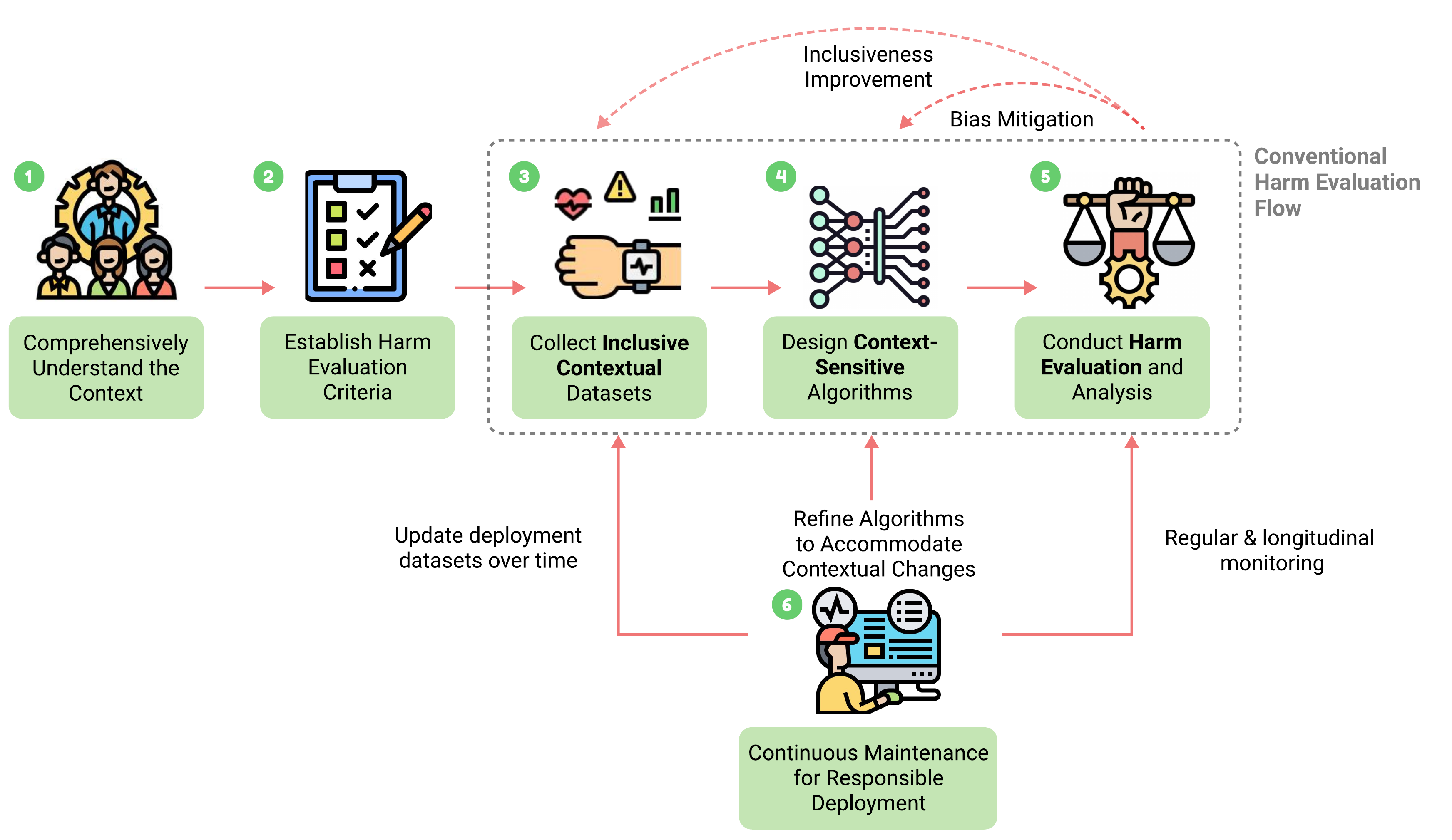}
    \caption{\textbf{Overview of the Framework for Evaluating and Mitigating Context-induced Harms in Behavioral Sensing}. Steps 3, 4, and 5 cover the conventional evaluation flow for identifying harms. }
    \label{fig:framework}
\end{figure}

In this section, we first present our developed framework for evaluating and mitigating the context-induced harms in behavioral sensing. We then compare our framework with the prevailing evaluation approach in behavioral sensing technology. Our framework, shown in Figure~\ref{fig:framework}, has six steps. 

\begin{itemize}
    \item \textbf{Step 1: Comprehensively understand the context.} In this initial phase of developing behavioral sensing technologies, a comprehensive understanding of the \textit{context} is necessary. This involves the awareness of users' diverse backgrounds and engaging with them to understand their specific needs, as well as considering the variety of situated settings (such as technology infrastructure and environmental conditions). 
    
     \item \textbf{Step 2: Establish criteria for evaluating harms, and make sure the bias is not attributed to random choice.} After obtaining a detailed understanding of the context, technology builders should select metrics that can effectively discern algorithmic variances in different contexts. To ensure that these differences are not attributed to random chance, technology builders should employ a rigorous quantitative method, predominantly a statistical analysis~\cite{yfantidou2023beyond}, for their assessment. The choice of these methods demands careful deliberation to address issues such as Type I errors (false positives) and variations from different groups.
     
    \item \textbf{Step 3: Evaluate whether the collected data includes inclusive and comprehensive contextual information.} This step involves evaluating whether the collected data adequately reflects the comprehensive context identified in Step 1. It is essential to ensure broad representation across a range of demographics and to be acutely aware of situational factors during data collection.
    
    \item \textbf{Step 4: Evaluate whether the technology uses context-sensitive algorithms.} This step focuses on assessing whether technology builders have engineered algorithms that are sensitive to potential harms and biases that may emerge from varying contexts. It often requires a careful approach to data selection for training and testing these algorithms. Moreover, technology builders should commit to a continuous cycle of refinement and improvement of these algorithms, especially when harm or bias is detected. 
   
    \item \textbf{Step 5: Conduct harm evaluation on the behavioral sensing system, combining user feedback and techniques to mitigate harms.} This step involves assessing whether technology builders comprehensively evaluate the behavioral sensing system to identify and address potential harms or biases. A crucial element of this evaluation is the integration of user feedback, which offers valuable insights into the technology's performance in real-world scenarios and its effects on various user groups. If any biases are detected, technology builders need to revisit Step 4. There, they must refine the algorithms by leveraging the insights gained from user feedback, ensuring the technology not only performs optimally but also responsibly.

    \item \textbf{Step 6: Continuous maintenance of data and algorithms for responsible deployment.} This step of evaluation is critical in the lifecycle of behavioral sensing technology. Once the technology is deployed, technology builders should continuously monitor and update both the data and algorithms and evaluate the model performance to ensure that the technology remains up-to-date and adapts to various contexts.
    
\end{itemize}



\section{Evaluating Existing Behavioral Sensing Technologies -- Two Evaluation Studies}\label{sec:case_study}

The objective of this section is two-fold. First, we aim to validate the feasibility of our proposed framework. Second, we strive to provide empirical evidence on the nuanced forms of identity-based harms and situation-based harms. To achieve these goals, we conduct post-hoc evaluations on two real-world behavioral well-being sensing technology studies, each within a different domain and involving a distinct ML task. The first study focuses on classifying students with depressive symptoms (Section~\ref{sec:case_study1}). The second study aims to regress students' engagement levels (Section~\ref{sec:case_study2}). 

For each evaluation, we start with a background section, delineating the real-world problem, the datasets used, as well as the ML task and algorithms chosen for our evaluation. We then describe our evaluation process (summarized in Table~\ref{tab:eval_methods}). This is followed by the evaluation results. Our evaluations mainly focus on two aspects.  
\begin{itemize}
    \item Evaluate the extent to which the steps proposed in our framework have been considered in previous efforts in the design and implementation of well-being sensing technologies.
    \item Identify the potential harms and biases these technologies might introduce to users, by performing a quantitative evaluation of those algorithms. 
\end{itemize} 

\begin{table}[t]
\sffamily
\caption{\textbf{Overview of Evaluation Criteria and Methods for Each Evaluation Study}. This table enumerates the various elements assessed in each evaluation study and lists the specific methods employed for evaluating the algorithms.}
\renewcommand{\arraystretch}{1.02}
   \resizebox{1\textwidth}{!}{
   \begin{tabular}{|L{0.33\textwidth}|L{0.5\textwidth}|L{0.5\textwidth}|}
\hline
  \cellcolor[HTML]{CCCCCC}\makecell[c]{\textbf{Framework Outline}} & \cellcolor[HTML]{CCCCCC}\makecell[c]{\textbf{\makecell[c]{Evaluation 1: Depression Detection}}} & \cellcolor[HTML]{CCCCCC}\makecell[c]{\textbf{\makecell[c]{Evaluation 2: Engagement Prediction}}} \\ \hline
Step 1: Comprehensively understand the context. &    \textbf{Identity-based harms}: gender, sexual orientation, race, immigration status, first-generation college student status, and disability status. \textbf{Situation-based bias}: device types and data collection time.  \textbf{Engagement with users}: understand their concerns and insights on depression detection sensing technologies.    &   \textbf{Identity-based harms}: gender, disability status, homeless youth, and religious minorities \textbf{Situation-based bias}: Temperature condition, location, and class group. \textbf{Engagement with users}:  understand their concerns and insights on student engagement prediction sensing technologies.       \\\hline
Step 2: Establish criteria for evaluating harms, and make sure the bias is not attributed to random choice.                     &     \textbf{Fairness metrics}: disparities in accuracy, false negative rate, and false positive rate. \textbf{Significance test}:   Mann-Whitney U test with Benjamini-Hochberg correction.      &     \textbf{Fairness metrics}:  disparities in mean squared error. \textbf{Significance test}: Linear mixed model.        \\ \hline
Step 3:  Evaluate whether the collected data includes inclusive and comprehensive contextual information.     &     \multicolumn{2}{l|}{\makecell[l]{Consider collecting data that could introduce identity-based harms and/or Situation-based bias \\ due to the contextual factors identified in Step 1.}}                     \\\hline
Step 4: Evaluate whether the technology uses context-sensitive algorithms.                               &     \multicolumn{2}{l|}{Ensure that the algorithms are aware of harms to users and can adapt to contextual changes.    }            \\ \hline
Step 5: Conduct harm evaluation on the behavioral sensing system, combining user feedback and techniques to mitigate harms.                             &          \multicolumn{2}{l|}{Assess algorithms for potential harms or biases and verify their mitigation with user feedback.}         \\ \hline
Step 6: Continuous maintenance of data and algorithms for responsible deployment.                                      &      \multicolumn{2}{l|}{Implement strategies and actions to regularly update and maintain the data and algorithms.}          \\ \hline
\end{tabular}}
\label{tab:eval_methods}
\end{table}

We provide detailed descriptions of the evaluation process for steps 1 and 2 within the respective sections of each evaluation study, as these steps are customized to each specific case. Additionally, to gain a deeper understanding of algorithmic harms, we conduct an experiment focusing on bias mitigation in each evaluation study. 

\subsection{Evaluation Study 1: Depression Detection}\label{sec:case_study1}

Research has been conducted using longitudinal passive sensing data from smartphones and wearable devices to predict and detect depression (\eg~\cite{wahle2016mobile,xu2021leveraging,wang2018tracking}). However, these studies often face challenges related to the limited access to datasets and algorithms, hindering reproducibility and transparency in the field. To address these issues, Xu \etal introduced GLOBEM~\cite{xu2023globem}, an open-sourced benchmark platform that includes implementations of nine depression detection algorithms and ten domain generalization algorithms. All depression detection algorithms focus on a common binary classification task: distinguishing whether users had at least mild depressive symptoms. They also released a four-year longitudinal passive sensing behavioral dataset from college students~\cite{xu2022globem}, aimed to highlight challenges in generalizing and reproducing longitudinal behavior models. In our evaluation study, we examine these depression detection algorithms through a lens focused on potential harms, employing the perspective provided by our proposed framework. Our goal is to assess whether the designs of these algorithms, or their implementations, have considered the steps outlined in our framework. If any of these steps were overlooked, what potential harms can we identify?

\textbf{\textit{Datasets}}. We chose the four datasets from the GLOBEM study~\cite{xu2022globem} for the evaluation of the depression detection algorithms. To facilitate analysis and comparison, these datasets were labeled chronologically according to the time of their collection (D1 to D4). These four datasets used in our evaluation consist of approximately 700 person-terms of data from around 500 unique participants who were enrolled in the same institution over 10 weeks during the Spring term between 2018 and 2021. These datasets include a wide range of passively sensed behavioral data, including sleep patterns, phone usage statistics, physical activity levels, and phone call records. The datasets also include a wide range of demographics, such as gender, race, first-generation status, immigration status, sexual orientation, and disability status. This data was continuously collected 24 hours per day from smartphones and Fitbits. Additionally, they also include self-reported depression data. We opted to use Beck Depression Inventory-II (BDI-II) scores~\cite{beck1996comparison}, which were collected once per person at the end of each term in each dataset, as the ground truth. 

\textbf{\textit{Depression Detection Algorithms}}.
We chose eight depression detection algorithms implemented by Xu \etal~\cite{xu2023globem}. These algorithms consist of a combination of support vector machine~\cite{canzian2015trajectories,farhan2016behavior,wahle2016mobile}, logistic regression~\cite{saeb2015mobile,wang2018tracking}, random forest~\cite{wahle2016mobile}, Adaboost~\cite{xu2019leveraging}, multi-task learning~\cite{lu2018joint}, and collaborative-filtering-based model~\cite{xu2021leveraging}. 
We excluded one algorithm in the implementation work developed by Chikersal \etal~\cite{chikersal2021detecting} from our evaluation study due to a significant disparity between our reproduced results and the reported results in the implementation work~\cite{xu2023globem} (shown in Table~\ref{tab:reproduction_results}).

\begin{table}[t]
\sffamily
\caption{\textbf{Reproduction Results}. Balanced accuracy of the nine depression prediction algorithms on four datasets: DS1 and DS2, which were used in prior work \cite{xu2023globem}, and DS3 and DS4, which are newly reported in this study. The comparison of algorithm performance on DS1 and DS2 ensures the reliability of our fairness evaluation. The $\Delta$ column represents the difference between our reproduced results and the previously reported results.}
\renewcommand{\arraystretch}{1.1}
   \resizebox{1\textwidth}{!}{\begin{tabular}{|l|rrr|rrr|r|r|}
\hline
\multicolumn{1}{|c|}{\multirow{2}{*}{\textbf{Algorithms}}}& \multicolumn{3}{c|}{\cellcolor[HTML]{CCCCCC}\textbf{DS1 (2018)}} & \multicolumn{3}{c|}{\cellcolor[HTML]{CCCCCC}\textbf{DS2 (2019)}} & \multicolumn{1}{c|}{\cellcolor[HTML]{CCCCCC}\textbf{DS3 (2020)}}         & \multicolumn{1}{c|}{\cellcolor[HTML]{CCCCCC}\textbf{DS4 (2021)}} \\ \cline{2-9} 
& \multicolumn{1}{l|}{Prior results} & \multicolumn{1}{l|}{Our results} & \multicolumn{1}{l|}{Diff ($\Delta$)} & \multicolumn{1}{l|}{Prior results} & \multicolumn{1}{l|}{Our results} & \multicolumn{1}{l|}{Diff ($\Delta$)} & \multicolumn{1}{l|}{Our results} & \multicolumn{1}{l|}{Our results} \\ \hline
Wahle \etal  \cite{wahle2016mobile}     & \multicolumn{1}{r|}{0.526}         & \multicolumn{1}{r|}{0.538}       & 0.012        & \multicolumn{1}{r|}{0.527}         & \multicolumn{1}{r|}{0.518}       & -0.009       & 0.514           & 0.514           \\ \hline
Saeb \etal \cite{saeb2015mobile}            & \multicolumn{1}{r|}{0.539}         & \multicolumn{1}{r|}{0.539}       & 0.000               & \multicolumn{1}{r|}{0.508}         & \multicolumn{1}{r|}{0.513}       & 0.005        & 0.588           & 0.500           \\ \hline
Farhan \etal \cite{farhan2016behavior}            & \multicolumn{1}{r|}{0.552}         & \multicolumn{1}{r|}{0.552}       & 0.000               & \multicolumn{1}{r|}{0.609}         & \multicolumn{1}{r|}{0.609}       & 0.000               & 0.563           & 0.609           \\ \hline
Canzian \etal \cite{canzian2015trajectories}           & \multicolumn{1}{r|}{0.559}         & \multicolumn{1}{r|}{0.538}       & -0.021       & \multicolumn{1}{r|}{0.516}         & \multicolumn{1}{r|}{0.516}       & 0.000               & 0.541           & 0.502           \\ \hline
Wang \etal  \cite{wang2018tracking} & \multicolumn{1}{r|}{0.566}         & \multicolumn{1}{r|}{0.565}       & -0.001       & \multicolumn{1}{r|}{0.500}         & \multicolumn{1}{r|}{0.500}       & 0.000               & 0.577           & 0.516           \\ \hline
Lu \etal \cite{lu2018joint} & \multicolumn{1}{r|}{0.574}         & \multicolumn{1}{r|}{0.574}       & 0.000               & \multicolumn{1}{r|}{0.558}         & \multicolumn{1}{r|}{0.558}       & 0.000               & 0.611           & 0.553           \\ \hline
Xu \etal - Interpretable \cite{xu2019leveraging} & \multicolumn{1}{r|}{0.722}         & \multicolumn{1}{r|}{0.688}       & -0.034       & \multicolumn{1}{r|}{0.623}         & \multicolumn{1}{r|}{0.667}       & 0.044        & 0.833           & 0.733           \\ \hline
Xu \etal - Personalized \cite{xu2021leveraging} & \multicolumn{1}{r|}{0.723}         & \multicolumn{1}{r|}{0.753}       & 0.030        & \multicolumn{1}{r|}{0.699}         & \multicolumn{1}{r|}{0.690}       & -0.009       & 0.791           & 0.686           \\ \hline
Chikersal \etal  (removed) \cite{chikersal2021detecting}       & \multicolumn{1}{r|}{\cellcolor[HTML]{EFEFEF}0.728}         & \multicolumn{1}{r|}{\cellcolor[HTML]{EFEFEF}0.618}       & \cellcolor[HTML]{EFEFEF}-0.110       & \multicolumn{1}{r|}{\cellcolor[HTML]{EFEFEF}0.776}         & \multicolumn{1}{r|}{\cellcolor[HTML]{EFEFEF}0.670}       & \cellcolor[HTML]{EFEFEF}-0.106       & \cellcolor[HTML]{EFEFEF}0.581           & \cellcolor[HTML]{EFEFEF}0.641           \\ \hline
\end{tabular}}
\label{tab:reproduction_results}
\end{table}

\subsubsection{Evaluation Methods and Results}\label{subsec:evaluation_methods} In this subsection, we elaborate on the decision-making processes involved in each step of our framework, followed by presenting our evaluation results.

\paragraph{\textbf{Step 1: Comprehensively understand the context}}\label{sec:identify_stakeholders} In the development of behavioral sensing technologies for depression detection, having a nuanced understanding of user diversity is crucial, as certain sub-populations exhibit higher depression rates. Studies indicate an increased prevalence of depression in specific demographics, such as women~\cite{albert2015depression}, first-generation college students~\cite{jenkins2013first}, immigrants~\cite{fung2010postpartum}, non-heterosexual individuals~\cite{zietsch2012shared}, racial minorities~\cite{breslau2006specifying}, and disability status~\cite{zhang2022impact}. These findings underscore the importance for depression detection technology builders to be aware of this context -- users' sensitive attributes -- to avoid societal biases and ensure equitable outcomes. Additionally, temporal factors and the type of devices used during data collection are crucial elements to consider. Research has indicated that depressive symptoms can fluctuate based on the time of day (\eg morning vs. evening)~\cite{putilov2018associations,chelminski1999analysis} and that user behaviors might be affected by the specific settings of their devices~\cite{church2015understanding}. Recognizing these contexts -- timing and device types during data collection -- is crucial. Such considerations enable technology builders to accurately model and predict depressive symptoms under different conditions, thereby ensuring the technology's adaptability and fairness in diverse data collection scenarios. Furthermore, engaging users early in the technology development process and incorporating their values and feedback is vital for increasing user acceptance and engagement~\cite{zhu2018value,friedman2013value}. This value-sensitive and human-centered approach ensures that the technology is not only technically sound but also resonates with the users' needs and preferences~\cite{mohr2017three,amershi2019guidelines}.

Building upon the above analysis, our evaluation focuses on assessing whether the designs of these depression detection algorithms take into account three critical aspects: the potential for identity-based harm, situation-based harm, and the extent of technology builders' engagement with users to understand their concerns about mental health sensing technologies (summarized in Table~\ref{tab:eval_methods}).

\paragraph{\textbf{Step 2: Establish criteria for evaluating harms, and make sure the bias is not attributed to random choice.}}\label{sec:define_eval} In this step, we evaluate whether prior work established criteria for evaluating potential harms. To facilitate us to thoroughly examine the potential harms introduced by these depression detection algorithms, we set two key evaluation criteria: classification fairness metrics and thresholds for quantifying differences and biases. In the following subsections, we elaborate on the decision-making process that guided our choices of these criteria. Additionally, we detail the experimental implementation in Appendix~\ref{appdx:example}, for the sake of transparency and to facilitate reproducibility.

\textit{\textbf{Criterion 1: Classification Fairness Metrics.}}\label{method_fairness_metrics}
We used three fairness metrics: disparity in accuracy, disparity in false negative rate, and disparity in false positive rate. These metrics were applied to assess algorithm performance across individuals with sensitive attributes and those without sensitive attributes. We intentionally chose not to adopt commonly used fairness metrics such as demographic parity (\eg~\cite{buet2022towards}), which aim to ensure equal treatment across different groups. This decision was based on prior research findings indicating that individuals with sensitive attributes are more likely to experience depressive symptoms (\eg~\cite{givens2007ethnicity,mcfadden2016health,lucero2012prevalence}). Using demographic parity, which aims for equal rates of predicted depressive symptoms across groups, could conflict with empirical evidence suggesting inherent disparities in depression prevalence. Our dataset analysis confirmed this, showing notably higher depression levels in certain sensitive groups (first-generation college students, immigrants, and non-male students) from 2018 and 2021\footnote{We performed a Mann-Whitney U test with the Benjamini-Hochberg correction for significance testing (more details are in Section \ref{method_stat_anlaysis}).} (see Figure \ref{fig_BDI2} in Appendix). This highlights the critical need for selecting fairness metrics that reflect real-world disparities.

\textit{\textbf{Criterion 2: Threshold for Quantifying Differences and Biases.}}\label{method_stat_anlaysis}
We further added a criterion: a threshold quantifying differences in algorithmic performances across various groups. We implemented this to mitigate the impact of random variations. For this purpose, we chose established statistical tests, specifically opting for a non-parametric approach, considering the non-normal distribution of the chosen datasets. We utilized the Mann-Whitney U test, a widely recognized method for comparing means between two independent samples, irrespective of their distribution~\cite{mann1947test, wilcoxon1992individual}. We further employed the Benjamini-Hochberg (B-H) correction method to manage the Type I error rate associated with multiple comparisons within the same dataset~\cite{benjamini1995controlling}. We set a stringent False Discovery Rate (FDR) threshold at 0.05~\cite{benjamini2005quantitative, glickman2014false}, ensuring that the rate of false positives is carefully controlled at 5\%.  

\paragraph{\textbf{Steps 3 to 6: Collect inclusive datasets including comprehensive contextual information; develop context-sensitive algorithms; evaluate the behavioral sensing, combining user feedback and techniques to mitigate harms; and continuous maintenance of data and algorithms for responsible deployment}}\label{sec:steps_3_6} In evaluating steps 3 to 6, we assessed how the existing depression detection sensing technology builders handled several crucial aspects. First, we looked at whether they took into account the potential for identity-based and situation-based harms during data collection, in line with the contextual factors highlighted in Step 1. Second, we examined the design of the algorithms to determine if they were conscious of potential harms or biases and if they could adapt to contextual changes. Third, we evaluated how these technologies mitigated harms identified in their algorithms, particularly focusing on the use of user feedback. Finally, we assessed whether there were effective strategies and actions in place for the regular updating and maintenance of the data and algorithms. 

Below, we present our evaluation findings of current depression detection algorithm designs, focusing on how well they align with our proposed framework. We then present the potential harms identified in these algorithms, based on the context identified in Step 1 and the criteria established in Step 2. This is followed by the experiment of bias mitigation and its results.

\begin{table}[htb!]
\caption{\textbf{Evaluation Results for Two Behavioral Sensing Technology Applications}. This table summarizes the assessment of eight depression detection algorithms (third to tenth rows) based on their original publications, and a re-implementation study (eleventh row). The final line evaluates a student engagement prediction model. Symbols indicate consideration levels: \cmark\ for full consideration, $\divideontimes$\ for partial consideration, and \xmark\ for no consideration.}\label{tab:overall_evaluation}
\Huge
\renewcommand{\arraystretch}{1.1}
   \resizebox{\textwidth}{!}{
\begin{tabular}{|l|c|c|c|c|c|c|}
\hline
\multicolumn{1}{|c|}{{\cellcolor[HTML]{CCCCCC}\textbf{\makecell[c]{Algorithm \\ Design / Implementation}}}} & 
\multicolumn{1}{c|}{{\cellcolor[HTML]{CCCCCC}\textbf{\makecell[c]{Comprehensively  \\ Understand Context}}}} & \multicolumn{1}{c|}{{\cellcolor[HTML]{CCCCCC}\textbf{\makecell[c]{Establish Harm \\ Evaluation Criteria}}}} & \multicolumn{1}{c|}{{\cellcolor[HTML]{CCCCCC}\textbf{\makecell[c]{Collect Inclusive \\ Contextual Datasets}}}} & \multicolumn{1}{c|}{{\cellcolor[HTML]{CCCCCC}\textbf{\makecell[c]{Design Context- \\ Sensitive Algorithms}}}} & \multicolumn{1}{c|}{{\cellcolor[HTML]{CCCCCC}\textbf{\makecell[c]{Harm Evaluation \\ and Mitigation}}}}  & \multicolumn{1}{c|}{{\cellcolor[HTML]{CCCCCC}\textbf{\makecell[c]{Continuous \\ Maintenance}}}} \\ \hline
\multicolumn{7}{|c|}{\cellcolor[HTML]{EFEFEF}\textbf{Evaluation Study 1: Depression Detection}} \\\hline
Wahle \etal & $\divideontimes$ & \xmark & \cmark &  \xmark & \xmark & \xmark   \\
Saeb \etal    & $\divideontimes$ & \xmark & \xmark & \xmark & \xmark  & \xmark  \\
Farhan \etal    & $\divideontimes$ & \xmark &  \cmark &  \xmark & \xmark  & \xmark  \\
Canzian \etal     & $\divideontimes$ & \xmark & \xmark &  \xmark & \xmark & \xmark  \\
Wang \etal & $\divideontimes$ & \xmark & \cmark &  \xmark & \xmark  & \xmark  \\
Lu \etal  & $\divideontimes$ & \xmark & \cmark &  \xmark & \xmark  & \xmark  \\
Xu \etal - Interpretable & $\divideontimes$ &  \xmark & \xmark & \xmark & \xmark  & \xmark  \\
Xu \etal - Personalized      & $\divideontimes$ & \xmark & \cmark & \xmark & \xmark & \xmark \\\hdashline
Xu \etal - Implementation      & $\divideontimes$ & \xmark & \cmark & \xmark & \xmark & \xmark \\\hline
\multicolumn{7}{|c|}{\cellcolor[HTML]{EFEFEF}\textbf{Evaluation Study 2: Engagement Prediction}} \\\hline
Gao \etal - En-gage & $\divideontimes$ & \xmark & \cmark & \xmark & \xmark & \xmark \\\hline
\end{tabular}}
\end{table}

\paragraph{\textbf{Evaluation Results}} 
In our review of nine papers related to the design and implementation of eight depression detection algorithms, we observed that none of the prior work discussed potential harms to users, neither of them engaged with users to better understand their needs. All prior work considered identity-based context, \ie sensitive attributes. However, consistent with previous sensing technology research, most studies only focused on two sensitive attributes: age~\cite{wahle2016mobile,farhan2016behavior,lu2018joint} and gender~\cite{farhan2016behavior,wang2018tracking,lu2018joint,xu2021leveraging,xu2023globem}. A few also considered race~\cite{farhan2016behavior,wang2018tracking,lu2018joint,xu2023globem}, but other sensitive attributes were largely overlooked. In terms of situated aspects, while most studies accounted for data collection time, consideration of device types was less common. Importantly, while these studies reported on this context information, many did not disclose the proportion of data pertaining to each, potentially leading to representative issues. More critically, none of the studies established criteria for evaluating potential harms, nor was there evidence of context-sensitive algorithm design or processes for harm evaluation and mitigation, particularly incorporating user feedback during the whole design process. Furthermore, there was a lack of strategies for the regular maintenance and updating of data and algorithms. These findings are summarized in Table~\ref{tab:overall_evaluation}, providing an overview of our evaluation results.

To assess the possibility of potential harms arising from a lack of context sensitivity in these depression detection algorithms, we carried out a quantitative analysis. Specifically, we leveraged the evaluation criteria defined in Step 2 (in Section~\ref{sec:define_eval}) and evaluated the eight depression detection algorithms on the five sensitive attributes identified in Step 1 (gender, first-generation college student status, immigration status, race, and sexual orientation). Note that, disability status was removed due to the small sample size. The results, detailed in Table \ref{tab_all_w_t_demo}, revealed several insights.

Firstly, we observed biases in all algorithms towards certain sensitive attributes, \ie their disparities in accuracy, false negative rates, and false positive rates (highlighted in red in Table~\ref{tab:overall_evaluation}). Notably, algorithms with higher balanced accuracy~\cite{xu2019leveraging,xu2022survey} tended to show fewer biases across these attributes when evaluated with the three fairness metrics. In particular, the algorithm \textbf{Xu\_interpretable} ~\cite{xu2019leveraging} did not exhibit bias in terms of accuracy and false positive rate disparities.

Another interesting finding was the reduced bias in all algorithms on DS3, the dataset collected at the start of the COVID-19 outbreak in 2020. This suggests that the significant impact of COVID-19 might have overshadowed other sensitive attributes, leading to this pattern of decreased bias. Additionally, we did not see a consistent pattern indicating which algorithms consistently demonstrated fair performance regarding the sensitive attributes.

\begin{table}[t]
\sffamily
\renewcommand{\arraystretch}{1.02}
\caption{\textbf{Algorithmic Harm Evaluation Results}. Results of algorithmic harms through the disparity in accuracy, the disparity in false negative rate, and the disparity in false positive rate (without incorporating demographic data into the training and testing process). The results are adjusted p-values by Benjamini-Hochberg correction after the Mann-Whitney U test. Significance is highlighted in red. Acc, Fnr, and Fpr are the abbreviations of the disparity in accuracy, the disparity in false negative rate, and the disparity in false positive rate.}
\resizebox{1\textwidth}{!}{\begin{tabular}{|l|l|ccc|ccc|ccc|ccc|}\hline
\multicolumn{1}{|c|}{\multirow{2}{*}{\textbf{Algorithms}}} &\multicolumn{1}{c|}{\multirow{2}{*}{\textbf{Sensitive Attributes}}} &
\multicolumn{3}{c|}{\cellcolor[HTML]{CCCCCC}\textbf{DS1 (2018)}} &
\multicolumn{3}{c|}{\cellcolor[HTML]{CCCCCC}\textbf{DS2 (2019)}} &
\multicolumn{3}{c|}{\cellcolor[HTML]{CCCCCC}\textbf{DS3 (2020)}} &
\multicolumn{3}{c|}{\cellcolor[HTML]{CCCCCC}\textbf{DS4 (2021)}} \\\cline{3-14}
& & \multicolumn{1}{c}{Acc}  & \multicolumn{1}{c}{Fnr}  & \multicolumn{1}{c|}{Fpr}  & \multicolumn{1}{c}{Acc}  & \multicolumn{1}{c}{Fnr}  & \multicolumn{1}{c|}{Fpr}  & \multicolumn{1}{c}{Acc}  & \multicolumn{1}{c}{Fnr}  & \multicolumn{1}{c|}{Fpr}  & \multicolumn{1}{c}{Acc}  & \multicolumn{1}{c}{Fnr}  & \multicolumn{1}{c|}{Fpr}  \\\hline

  \multirow{5}{*}{Wahle \etal \cite{wahle2016mobile}} & First-gen College Student & 0.020 & 0.030 & 0.030 &  \cellcolor[HTML]{F4CCCC}0.010 & 0.050 & \cellcolor[HTML]{F4CCCC}0.010 & 0.020 & 0.030 & 0.040 & 0.010 & 0.050 & 0.020 \\
  & Gender & 0.030 & 0.030 & 0.030 & \cellcolor[HTML]{F4CCCC}0.020 & 0.030 & 0.030 & 0.050 & 0.010 & 0.020 & 0.050 & 0.030 & 0.050 \\
  & Immigration Status & 0.040 & 0.030 & 0.030 & 0.040 & 0.040 & 0.050 & 0.010 & 0.020 & 0.040 & 0.040 & 0.010 & 0.030 \\
  & Race & \cellcolor[HTML]{F4CCCC}0.010 & 0.030 & 0.030 & 0.030 & 0.010 & 0.020 & 0.030 & 0.050 & 0.030 & 0.020 & 0.020 & 0.010 \\
  & Sexual Orientation & 0.050 & 0.030 & 0.030 & 0.050 & 0.020 & 0.040 &  0.040 & 0.040 & 0.010 & 0.030 & 0.040 & 0.040 \\
\hline

\multirow{5}{*}{Saeb \etal \cite{saeb2015mobile}} & First-gen College Student & \cellcolor[HTML]{F4CCCC}0.010 & 0.030 & 0.010 & \cellcolor[HTML]{F4CCCC}0.020 & 0.030 & 0.030 & 0.010 & 0.050 & 0.050 & \cellcolor[HTML]{F4CCCC}0.020 & 0.030 & 0.050\\
  & Gender & 0.050 & 0.040 &0.050 & \cellcolor[HTML]{F4CCCC}0.030 & 0.030 & 0.030 & 0.020 & 0.030 & \cellcolor[HTML]{F4CCCC}0.020 & 0.040  & 0.040 & 0.030 \\
  & Immigration Status & \cellcolor[HTML]{F4CCCC}0.020 &\cellcolor[HTML]{F4CCCC}0.010 & 0.040 &  \cellcolor[HTML]{F4CCCC}0.010 & 0.030 & 0.030 & 0.040 & 0.040 & 0.030 & 0.050 & 0.050 & 0.020\\
  & Race & \cellcolor[HTML]{F4CCCC}0.030 & 0.050 & 0.020 & \cellcolor[HTML]{F4CCCC}0.040 & 0.030 & 0.030 & 0.050 & 0.010 & \cellcolor[HTML]{F4CCCC}0.010 & 0.030  & 0.020 & 0.010\\
  & Sexual Orientation & 0.040 & \cellcolor[HTML]{F4CCCC}0.020 & 0.030 & 0.050 & 0.030 & 0.030 &  0.030 & 0.020 & 0.040 & \cellcolor[HTML]{F4CCCC}0.010 & \cellcolor[HTML]{F4CCCC}0.010 & 0.040 \\
\hline

\multirow{5}{*}{Farhan \etal \cite{farhan2016behavior}} & First-gen College Student & 0.030 & 0.020 & 0.040 & 0.020 & 0.030 & 0.040 & 0.030 & 0.040 & 0.040 &  0.030 & 0.010 & 0.010 \\
  & Gender & 0.020 & 0.030 & 0.030 &  \cellcolor[HTML]{F4CCCC}0.010 & 0.020 & 0.010 & 0.040 & 0.030 & 0.050 & 0.010 & 0.040 & 0.030 \\
  & Immigration Status & 0.040 & 0.040 & 0.020 & 0.040 & \cellcolor[HTML]{F4CCCC}0.010 & 0.030 & 0.050 & 0.010 & 0.030 & 0.050 & 0.050 & 0.050 \\
  & Race & 0.050 & 0.050 & 0.010 & 0.050 & 0.050 & 0.050 & 0.010 & 0.020 & \cellcolor[HTML]{F4CCCC}0.010 & 0.040 & 0.020 & 0.040\\
  & Sexual Orientation & 0.010 & 0.010 & 0.050 & 0.030 & 0.040 & 0.020 & 0.020 & 0.050 & \cellcolor[HTML]{F4CCCC}0.020 & 0.020 & 0.030 & 0.020 \\
\hline

\multirow{5}{*}{Canzian \etal \cite{canzian2015trajectories}} & First-gen College Student & 0.020 & 0.030 & 0.030 & \cellcolor[HTML]{F4CCCC}0.020 & 0.010 & \cellcolor[HTML]{F4CCCC}0.020 & 0.020 & 0.020 & 0.030 & \cellcolor[HTML]{F4CCCC}0.010 & 0.030 & 0.020 \\
  & Gender & 0.030 & 0.030 & 0.030 & \cellcolor[HTML]{F4CCCC}0.010 & 0.020 & \cellcolor[HTML]{F4CCCC}0.010 & 0.020 & 0.030 & 0.020 & 0.030  &0.020 & 0.030 \\
  & Immigration Status & 0.040 & 0.030 & 0.030 & \cellcolor[HTML]{F4CCCC}0.030 & 0.050 & 0.050 & 0.050 & 0.040 & 0.040 & 0.050 & 0.040 & 0.050 \\
  & Race & \cellcolor[HTML]{F4CCCC}0.010 & 0.010 & 0.030 &  0.050 & 0.030 & 0.040 & 0.010 & 0.020 & \cellcolor[HTML]{F4CCCC}0.010 & 0.040 & 0.030 & 0.040 \\
  & Sexual Orientation & 0.050 & 0.030 & 0.030 & 0.040 & 0.030 & 0.040 & 0.040 & 0.050 & 0.050 & \cellcolor[HTML]{F4CCCC}0.020 & 0.050 & \cellcolor[HTML]{F4CCCC}0.010 \\
\hline
 
\multirow{5}{*}{Wang \etal \cite{wang2018tracking}} & First-gen College Student  & 0.020 & 0.050 & 0.040 & \cellcolor[HTML]{F4CCCC}0.020 & 0.030 & 0.030 & 0.040 & 0.030 & 0.050 & \cellcolor[HTML]{F4CCCC}0.020 & 0.040 & 0.030 \\
  & Gender & 0.040 & 0.040 & 0.030 & \cellcolor[HTML]{F4CCCC}0.030 & 0.030 & 0.030 & 0.050 & 0.040 & 0.020 & 0.040 & 0.020 & 0.040 \\
  & Immigration Status & \cellcolor[HTML]{F4CCCC}0.010 & 0.020 & 0.010 & \cellcolor[HTML]{F4CCCC}0.010 & 0.030 & 0.030 & 0.010 & 0.010 & 0.030 & 0.050 & 0.010 & 0.050 \\
  & Race & 0.030 & 0.010 & 0.020 & \cellcolor[HTML]{F4CCCC}0.040 & 0.030 & 0.030 & 0.020 & 0.020 & 0.040 &0.030 & 0.030 & 0.020 \\
  & Sexual Orientation & 0.050 & 0.030 & 0.050 & 0.050 & 0.030 & 0.030 & 0.030 & 0.050 & 0.010 & \cellcolor[HTML]{F4CCCC}0.010 & 0.050 & \cellcolor[HTML]{F4CCCC}0.010 \\
\hline

\multirow{5}{*}{Lu \etal \cite{lu2018joint}} & First-gen College Student & \cellcolor[HTML]{F4CCCC}0.010 & \cellcolor[HTML]{F4CCCC}0.020 & 0.030 & 0.010 & 0.010 & 0.050 & 0.040 & 0.040 & 0.040 & 0.050 & 0.040 & 0.030 \\
& Gender & 0.050 & 0.030 & 0.020 & 0.040 & 0.020 & 0.020 & 0.030 & 0.040 & 0.050 & 0.020 & 0.030 & 0.020 \\
& Immigration Status & 0.040 & 0.050 & 0.040 & 0.050 & 0.030 & 0.040 & 0.050 & 0.010 & 0.020 & 0.010 & 0.050 & \cellcolor[HTML]{F4CCCC}0.010 \\
& Race & 0.020 & \cellcolor[HTML]{F4CCCC}0.010 & 0.010 & 0.030 & 0.040 & 0.030 & 0.010 & 0.020 & 0.030 & 0.040 & 0.010 & 0.050 \\
& Sexual Orientation & 0.030 & 0.040 & 0.030 & 0.020 & 0.050 & 0.050 &  0.020 & 0.030 & 0.050 & 0.030 & 0.020 & 0.020 \\
\hline
 
 \multirow{5}{*}{Xu\_interpretable \etal \cite{xu2019leveraging}} & First-gen College Student & 0.040 & 0.020 & 0.050 & 0.010 & 0.030 & 0.010 & 0.010 & \cellcolor[HTML]{F4CCCC}0.010 & 0.020 & 0.030 & 0.010 & 0.030 \\
  & Gender & 0.020 & 0.040 & 0.040 & 0.020 & 0.010 & 0.020 & 0.040 & 0.030 & 0.040 & 0.010 & 0.050 & 0.010 \\
  & Immigration Status & 0.030 & 0.010 & 0.020 & 0.050 & 0.050 & 0.030 & 0.030 & 0.020 & 0.030 & 0.050 & 0.030 & 0.040 \\
  & Race & 0.010 & 0.030 & 0.010 & 0.030 & 0.020 & 0.040 & 0.020 & 0.040 & 0.010 & 0.040 & 0.020 & 0.050 \\
  & Sexual Orientation & 0.050 & 0.050 & 0.030 & 0.040 & 0.040 & 0.050 & 0.050 & 0.050 & 0.050 & 0.020  & 0.040 & 0.020 \\
\hline
 
 \multirow{5}{*}{Xu\_personalized \etal \cite{xu2021leveraging}} & First-gen College Student & 0.030 & 0.040 & 0.020 & 0.030 & 0.030 & 0.040 & 0.030 & 0.040 & 0.020 & 0.020 & 0.020 & 0.010 \\
  & Gender & 0.010 & 0.050 & 0.040 & 0.040 & 0.050 & 0.030 & 0.050 & 0.010 & 0.010 & 0.050 & 0.030 & 0.020 \\
  & Immigration Status & 0.050 & 0.020 & 0.050 & 0.020 & 0.040 & 0.020 & 0.010 & 0.030 & 0.020 & 0.040 & 0.050 & 0.030 \\
  & Race & 0.020 & 0.010 & 0.030 & \cellcolor[HTML]{F4CCCC}0.010 & 0.020 & \cellcolor[HTML]{F4CCCC}0.010 & 0.020 & 0.020 & 0.050 & 0.030  & 0.040 &0.040 \\
  & Sexual Orientation & 0.040 & 0.030 & 0.010 & 0.050 & 0.010 & 0.050 & 0.040 & 0.050 & 0.040 & 0.010 & \cellcolor[HTML]{F4CCCC}0.010 & 0.050 \\
\hline
 
\end{tabular}}
\label{tab_all_w_t_demo}
\end{table} 

\paragraph{\textbf{Additional Experiment on Bias Mitigation}}\label{sec:bias_mitigation}
Recognizing the presence of biases in the eight algorithms, we took steps to mitigate these algorithmic biases.
Our approach involved an in-processing technique, where sensitive attributes were incorporated into both the training and testing phases~\cite{yfantidou2023beyond,wan2023processing}. This method allows algorithms to understand and learn from the relationships between sensitive attributes and the target variable (\ie BDI-II scores). Previous research indicates that such an approach can help diminish discriminatory patterns present in the data, thereby enhancing the fairness of the models across diverse groups.~\cite{yfantidou2023beyond,wan2023processing}. Our goal in this experiment was not to develop specific fair algorithms or mitigation techniques but rather to demonstrate a method for reducing bias and obtaining new insights.

As an example, we selected \textbf{Xu\_interpretable} algorithm~\cite{xu2019leveraging} due to its relatively high detection performance in depression detection (shown in Table~\ref{tab:reproduction_results}) and its relatively low level of disparities across three fairness metrics in the four datasets and five sensitive attributes. We focused on mitigating bias related to first-generation college student status, a sensitive attribute where this algorithm showed bias in terms of disparity of false negative rate.  We integrated this attribute into both the training and testing phases of the algorithm and re-evaluated the algorithm's performance. 

\begin{table}[htb!]
\sffamily
\renewcommand{\arraystretch}{1.02}
\caption{\textbf{Results After Implementing Bias Mitigation Techniques}. Example of algorithmic fairness evaluation results of Xu\_Interpretable~\cite{xu2019leveraging} through the three fairness metrics (with incorporating first-generation college student status into the training and testing process). The first row in each sub-table showcases the result of our evaluation pertaining to the first-generation college student status, subsequent to the incorporation of this sensitive attribute into both the training and testing phases. In the context of q-values obtained before and after the inclusion of first-generation college student status, $\spadesuit$ represents the former, while $\blacksquare$ represents the latter. Acc, Fnr, and Fpr are the abbreviations of the disparity in accuracy, the disparity in false negative rate, and the disparity in false positive rate.}
\resizebox{1\textwidth}{!}
{\begin{tabular}{|l|l|rr|rr|rr|rr|}
\hline
\multicolumn{1}{|c|}{\multirow{2}{*}{\textbf{Fairness Metric}}} & \multicolumn{1}{c|}{\multirow{2}{*}{\textbf{Sensitive Attribute}}} & \multicolumn{2}{c|}{\cellcolor[HTML]{CCCCCC}\textbf{DS1 (2018)}} & \multicolumn{2}{c|}{\cellcolor[HTML]{CCCCCC}\textbf{DS2 (2019)}} & \multicolumn{2}{c|}{\cellcolor[HTML]{CCCCCC}\textbf{DS3 (2020)}} & \multicolumn{2}{c|}{\cellcolor[HTML]{CCCCCC}\textbf{DS4 (2021)}} \\ \cline{3-10} 
\multicolumn{1}{|c|}{}& \multicolumn{1}{c|}{}    & \multicolumn{1}{c}{q value $\spadesuit$}  & \multicolumn{1}{c|}{q value $\blacksquare$}  & \multicolumn{1}{c}{q value $\spadesuit$}  & \multicolumn{1}{c|}{q value $\blacksquare$}  & \multicolumn{1}{c}{q value $\spadesuit$}  & \multicolumn{1}{c|}{q value $\blacksquare$}  & \multicolumn{1}{c}{q value $\spadesuit$}  & \multicolumn{1}{c|}{q value $\blacksquare$}   \\ \hline
\multicolumn{1}{|l|}{\multirow{5}{*}{Disparity in Acc}}    & \textbf{First-gen College Student}& \textbf{0.040} & \textbf{0.030}& \textbf{0.010} & \textbf{0.020}& \textbf{0.010} & \textbf{0.040}& \textbf{0.030} & \textbf{0.030} \\
\multicolumn{1}{|c|}{}& Gender& 0.020 & 0.010& 0.020 & 0.010& 0.040 & 0.020& 0.010 & \cellcolor[HTML]{F4CCCC}*0.010\\
\multicolumn{1}{|c|}{}& Immigration Status  & 0.030 & 0.050 & 0.050 & 0.050 & 0.030 & 0.030 & 0.050 & 0.040 \\
\multicolumn{1}{|c|}{}& Race& 0.010 & 0.040 & 0.030 & 0.040 & 0.020 & 0.010 & 0.040 & 0.020 \\
\multicolumn{1}{|c|}{}& Sexual Orientation  & 0.050 & 0.020 & 0.040 & 0.030 & 0.050 & 0.050 & 0.020 & 0.050 \\ \hline
\multirow{5}{*}{Disparity in Fnr} & \textbf{First-gen College Student}& \textbf{0.020} & \textbf{0.040}& \textbf{0.030} & \textbf{0.020}& \cellcolor[HTML]{F4CCCC}\textbf{*0.010} & \textbf{0.030}& \textbf{0.010} &  \textbf{0.020} \\
 & Gender& 0.040 & 0.010& 0.010 & 0.030& 0.030 & 0.050& 0.050 & 0.050\\
 & Immigration Status  & 0.010 & 0.030& 0.050 & 0.050& 0.020 & 0.040& 0.030 & 0.040\\
 & Race& 0.030 & 0.020& 0.020 & 0.040& 0.040 & 0.020& 0.020 & 0.030\\
 & Sexual Orientation  & 0.050 & 0.050& 0.040 & 0.010& 0.050 & 0.010& 0.040 & 0.010   \\\hline
\multirow{5}{*}{Disparity in Fpr} & \textbf{First-gen College Student}& \textbf{0.050} & \textbf{0.030}& \textbf{0.010} & \textbf{0.020}& \textbf{0.020} & \textbf{0.040}& \textbf{0.030} & \textbf{0.030} \\
 & Gender& 0.040 & 0.010 & 0.020 & 0.010 & 0.040 & 0.020& 0.010 & \cellcolor[HTML]{F4CCCC}*0.010   \\
 & Immigration Status  & 0.020 & 0.020 & 0.030 & 0.040 & 0.030 & 0.050 & 0.040 & 0.050 \\
 & Race& 0.010 & 0.050 & 0.040 & 0.030 & 0.010 & 0.010 & 0.050 & 0.020 \\
 & Sexual Orientation  & 0.030 & 0.035& 0.040 & 0.038& 0.050 & 0.027& 0.030 & 0.040  \\\hline 
\end{tabular}}
\label{tab:xu_interpretable_first-gen}
\end{table}

Our evaluation result, as detailed in Table \ref{tab:xu_interpretable_first-gen}, demonstrates the effectiveness of including the status of being a first-generation college student in the training and testing phases to reduce algorithmic harms. This method led to a fair treatment of this particular sensitive attribute across all datasets, evaluated using three different fairness metrics. However, it is worth noting that while this approach improved fairness for first-generation college student status, it seemed to adversely impact fairness concerning other sensitive attributes such as sexual orientation and gender. 
A more comprehensive discussion of such trade-offs is described in Section \ref{sec:discussion}.

\subsection{Evaluation Study 2: Student Engagement Prediction}\label{sec:case_study2}

\subsubsection{Background}
In recent years, addressing the growing concerns of poor academic performance and student disinterest has led to a heightened interest in understanding student engagement, emotions, and daily behavior. This shift has coincided with significant advances in sensing technology, paving the way for novel methods to unobtrusively monitor and analyze student behavior and mental well-being in educational settings. A significant milestone in this domain is the introduction of the \textit{En-Gage} dataset by Gao \etal~\cite{gao2022understanding}. This dataset, available at PhysioNet \footnote{The dataset download link: \url{https://physionet.org/content/in-gauge-and-en-gage/1.0.0/}}, is distinguished as the largest and most diverse dataset in environmental and affect sensing within the educational field, offering unparalleled insights into student engagement patterns and classroom dynamics through a diverse array of sensing technologies.

\textit{Dataset}. The En-Gage dataset includes a four-week cross-sectional study involving 23 Year-10 students (15–17 years old, 13 female and 10 male) and 6 teachers (33–62 years old, four female and two male) in a mixed-gender K12 private school. It utilizes wearable sensors to collect physiological data and daily surveys to gather information on the participants’ thermal comfort (the comfort level of students regarding the perceived temperature at the time), learning engagement, seating locations, and emotions during school hours. An initial online survey was conducted to obtain participants' background information, including age, gender, general thermal comfort, and class groups. The dataset reflects the students' organization into different groups (Form group, Math group, and Language group), aiding in tracking their classroom locations. To clarify, students are typically enrolled in courses based on their form group division, except for math courses which are determined by their math group division, and language courses which are determined by their language group division.

Throughout the study, the participants were asked to wear \textit{Empatica E4} wristbands \cite{mccarthy2016validation} during school time, which capture 3-axis accelerometer readings, electrodermal activity, photoplethysmography (PPG), and skin temperature. They were also asked to complete online surveys three times a day, posted after certain classes. These surveys capture detailed insights into participants' behavioral, emotional, and cognitive engagement, as well as their emotions, thermal comfort and seating locations \cite{gao2022individual}. In total, the dataset comprises 291 survey responses and 1415.56 hours of physiological data from all participants. 

\textit{Engagement Prediction Models.} We chose the engagement regression model, LightGBM Regressors~\cite{nemeth2019comparison}, developed by Gao \etal~\cite{gao2020n}. The regression model is designed to predict student engagement across three dimensions: \textit{emotional}, \textit{cognitive}, and \textit{behavioral} engagement. Emotional engagement evaluates their feelings of belonging and emotional reaction to the educational environment, cognitive engagement assesses their effort to understand complex ideas and skills, and behavioral engagement looks at students' participation in academic and extracurricular activities. The 1 to 5 Likert scale was used for scoring engagement levels, where 1 represents low and 5 high engagement. To predict these multidimensional scores, a variety of features were extracted, including data from wearable devices and weather stations. It is worth noting that, data such as gender, thermal comfort, and class groups, were not used for the engagement prediction.

\subsubsection{Evaluation Methods and Results.}
In this subsection, following the approach used in Subsection~\ref{subsec:evaluation_methods}, we first explain the decision-making process for each step of our framework, followed by the results of our evaluation.

\paragraph{\textbf{Step 1: Comprehensively understand the context}} Previous research has highlighted that marginalized groups, including women of color, students with disabilities, homeless youth, and religious minority students, often face feelings of alienation and isolation, which can significantly affect their learning engagement~\cite{quaye2019student,sjogren2021complexities}. Studies have also pointed out that variables like the perceived temperature and the timing of data collection can impact student engagement~\cite{garrosa2017curiosity,morrison2020quantifying}. Moreover, the role of social learning spaces, derived from engaging with student participants, has been recognized as a factor contributing to enhanced engagement~\cite{matthews2011social}. 

Given these insights, our evaluation of student engagement prediction technology focused on whether its design took these contextual factors into account and aligned with the values and experiences of users. An overview of contextual factors we evaluated can be found in Table~\ref{tab:eval_methods}.

\paragraph{\textbf{Step 2: Establish criteria for evaluating harms, and make sure the bias is not attributed to random choice}} Given the regression-based prediction task, we used the disparity in Mean Squared Error (MSE) as our primary fairness metric to identify biases in model performance. MSE, the average of squared discrepancies between predicted and actual values, is widely recognized for assessing regression model accuracy~\cite{wang2009mean}. Additionally, to discern if biases were systematic or due to random variation, and considering repeated measurements from individuals, we adopted a linear mixed model method ~\cite{laird1982random}. This approach involved calculating residuals (differences between actual values and predictions) across various engagement prediction tasks. Subsequently, we utilized a linear mixed model, executed in Python, to examine whether these residuals significantly varied among different groups (\eg gender and thermal comfort). This statistical method is beneficial for its ability to account for both within-group and between-group variations in the data, thereby offering a deeper insight into the biases present in model performance. 

\paragraph{\textbf{Steps 3 to 6}} Our approach to decision-making and evaluation for Steps 3 to 6 in this evaluation study mirrors the method we employed in the first evaluation study, detailed in Section~\ref{sec:steps_3_6}. A comprehensive overview of the criteria and methods used in these steps can also be found in Table~\ref{tab:eval_methods}.

\textbf{\textit{Evaluation Results}}. In line with the results from our first evaluation study, our examination of relevant papers in this study~\cite{gao2020n,gao2022understanding} indicates that researchers did not engage with the users to understand their needs or considered potential harms to users, and only very limited contextual factors were considered. These factors included gender, thermal comfort at the time of data collection, and information about the courses and classrooms that participants were involved in prior to data collection. Additionally, a key observation is that while this contextual data was considered during the data collection phase, it was not actively incorporated into the training and testing phases of their algorithms. Moreover, the researchers did not address the potential harms of their algorithms. They did not establish criteria for evaluating such harms or implement techniques, including student feedback, to mitigate potential biases. Additionally, there was no evidence of strategies for regular maintenance and updates of the data and algorithms.

\begin{table}[t]
\renewcommand{\arraystretch}{1.05}
\caption{\textbf{Results of Linear Mixed Models Analysis}. This table displays the results from linear mixed models, focusing on identifying the significance of differences in regression models across diverse contexts within different engagement prediction tasks. Levels of significance are denoted as follows: * for $p<0.05$, ** for $p<0.01$, and *** for $p<0.001$. For each contextual factor, one group is designated as the reference (or baseline) category, for example, the Female group in Gender. The ``Interpret'' represents the average effect for the reference group when all other predictors are held at their reference level (for categorical variables).}
\resizebox{1\textwidth}{!}
{\begin{tabular}{|l|l|rrr|rrr|rrr|}
\hline
\multicolumn{1}{|c|}{\multirow{2}{*}{\textbf{Contextual Factors}}} & \multicolumn{1}{c|}{\multirow{2}{*}{\textbf{Model Variables}}}& \multicolumn{3}{c|}{\cellcolor[HTML]{CCCCCC}\textbf{Emotional Engagement}} & \multicolumn{3}{c|}{\cellcolor[HTML]{CCCCCC}\textbf{Cognitive Engagement}} & \multicolumn{3}{c|}{\cellcolor[HTML]{CCCCCC}\textbf{Behavioral Engagement}}\\\cline{3-11} 
\multicolumn{1}{|c|}{}& \multicolumn{1}{c|}{}  &  \makecell[c]{Coef.} & \makecell[c]{Std. Error} &  \makecell[c]{P>|z|} & \makecell[c]{Coef.} & \makecell[c]{Std. Error} &  \makecell[c]{P>|z|} & \makecell[c]{Coef.} & \makecell[c]{Std. Error} &  \makecell[c]{P>|z|}  \\
\hline
\multirow{3}{*}{Gender}& Intercept & 0.006  &   0.119 & 0.962 & -0.055  &   0.121 & 0.651 & -0.033 & 0.105 & 0.753 \\
& groups [T.Male] & -0.025  &   0.179 & 0.891  & 0.080  &   0.182  & 0.659 & -0.038 & 0.157 & 0.810 \\
& Group Var & 0.121  &  0.072 & & 0.120  &  0.070  & & 0.076 & 0.044 & \\\hline
\multirow{4}{*}{Thermal Comfort}& Intercept &  -0.140 &  0.116  & 0.225   &  -0.113  &   0.118 & 0.338 & -0.136 & 0.113 & 0.232 \\
& groups [T.No change] & 0.286 & 0.112 &  \cellcolor[HTML]{F4CCCC}*\textbf{0.011}  & 0.181  &   0.119  & 0.130 &0.215  & 0.117 & 0.067 \\
& groups [T.Warmer] & -0.117 & 0.150 & 0.436 & -0.013  &   0.160 & 0.937 & -0.190 & 0.156 & 0.225\\
& Group Var & 0.112 & 0.067  & & 0.098  &  0.054 & & 
 0.081 & 0.052  & \\\hline
\multirow{5}{*}{Language Group }& Intercept & 0.001  &  0.094 & 0.991 & 0.101  &   0.114  & 0.377 & 0.021 &  0.102 & 0.836 \\
& groups [T.Room 41] & 0.721   & 0.244 &  \cellcolor[HTML]{F4CCCC}**\textbf{0.003}  & -0.492  &   0.299 &  0.100 & 0.327 &0.264  & 0.215\\
& groups [T.Room 43] & -0.094   &  0.184  & 0.611 & -0.213 &    0.219  & 0.331  & -0.248 & 0.198 & 0.210 \\
& groups [T.Room 68] & -0.351  &  0.198 & 0.076 &  -0.181 &    0.241  & 0.452   & -0.324 & 0.214 & 0.129 \\
& Group Var & 0.057  &  0.045 & & 0.097  &  0.070 & & 0.068 & 0.052 & \\\hline
\multirow{3}{*}{Math Group}& Intercept & 0.194 &    0.166 & 0.242 & -0.314 &    0.155 &  \cellcolor[HTML]{F4CCCC}*\textbf{0.043} &0.125 & 0.153 & 0.414 \\
& groups [T.Room 41] & -0.292  &   0.215 & 0.175 & 0.330 &    0.201  & 0.101 & -0.335 & 0.198 & 0.091 \\
& groups [T.Room 43] &  -0.251 &    0.223 & 0.261 & 0.500  &   0.209 & 0.017  & -0.131 & 0.205 & 0.524 \\
& Group Var & 0.111  &  0.070 & & 0.086  &  0.05 & &0.082  & 0.055 & \\\hline
\multirow{9}{*}{Course}& Intercept & -0.285  &   0.237 & 0.228 &  -0.376  &   0.241& 0.119 &-0.022 & 0.237 & 0.926 \\
& groups [T.English] &  0.488 &    0.241 & \cellcolor[HTML]{F4CCCC}*\textbf{0.043} & 0.396  &   0.246  & 0.107 & 0.290 & 0.246 & 0.239 \\
& groups [T.Health] &  0.400  &   0.353 & 0.257 & 0.395   &  0.360  & 0.273 & -0.086 & 0.360 & 0.811 \\
& groups [T.Language] & 0.075 &    0.255 & 0.769 &  -0.148  &  0.260 & 0.569 & -0.400 & 0.260 & 0.124 \\
& groups [T.Maths] &  0.274 &    0.240  & 0.253 & 0.530  &   0.245 &  \cellcolor[HTML]{F4CCCC}*\textbf{0.030} & 0.035 & 0.245 & 0.885 \\
& groups [T.PE] &  0.485  &   0.356 & 0.173 & 0.558  &    0.363 &   0.125  & 0.226 & 0.363 & 0.532 \\
& groups [T.Politics] &  0.174 &    0.251  & 0.489 & 0.212  &   0.256 &  0.407 & -0.343 & 0.257 & 0.182 \\
& groups [T.Science] &  0.240 &     0.262 &   0.360 & 0.634 &   0.267 & \cellcolor[HTML]{F4CCCC}*\textbf{0.018}  & -0.085 & 0.267 & 0.749 \\
& Group Var &  0.128  &  0.085 & & 0.126  &  0.073 & & 0.084 & 0.054 & \\\hline

\end{tabular}}
\label{tab:case_study2_eval_results}
\end{table}

\begin{table}[t]
    \renewcommand{\arraystretch}{1.02}
    \centering
    \caption{\textbf{Overview of Basic Statistics.} MSE refers to the Mean Squared Error, indicating the average of the squares of the errors. 'Residual' denotes the difference between the ground truth and prediction. MR represents the Mean Residual, which is the average of residuals within each group. ``Ind'' and ``Obs'' stand for individuals and observations, respectively. 
    }
    \resizebox{0.7\textwidth}{!}{\begin{tabular}{|l|l|r|rr|rr|rr|}
    \hline
    \multicolumn{1}{|c|}{\multirow{2}{*}{\textbf{Context Factors}}} & \multicolumn{1}{c|}{\multirow{2}{*}{\textbf{Groups}}} & \multicolumn{1}{c|}{\multirow{2}{*}{\textbf{\makecell[c]{Counts \\ (Ind/Obs)}}}} & \multicolumn{2}{c|}{\cellcolor[HTML]{CCCCCC}\textbf{\makecell[c]{Emotional \\ Engagement}}} & \multicolumn{2}{c|}{\cellcolor[HTML]{CCCCCC}\textbf{\makecell[c]{Cognitive \\ Engagement}}} & \multicolumn{2}{c|}{\cellcolor[HTML]{CCCCCC}\textbf{\makecell[c]{Behavioral \\ Engagement}}} \\\cline{4-9} 
    \multicolumn{1}{|c|}{} & \multicolumn{1}{c|}{} & \multicolumn{1}{c|}{} & MSE & MR & MSE & MR & MSE & MR \\\hline
    \multirow{2}{*}{Gender} & Female & 13/149 & 0.708 &  -0.014 & 0.711 & -0.005 & 0.800 & -0.023\\
    & Male & 10/142 & 0.693 & 0.033 & 0.822 & -0.004 & 0.674 & 0.002 \\\hline
    \multirow{3}{*}{Thermal Comfort} & No Change & 22/163 & 0.631 & 0.158 & 0.774 & 0.069 & 0.687 & 0.129 \\
    & Cooler & 20/77 & 0.822 & -0.140 & 0.755 & -0.135 & 0.730 & -0.115 \\ 
    & Warmer & 14/51 & 0.742 & -0.242 & 0.751 & -0.045 &  0.915 & -0.300 \\\hline
    \multirow{4}{*}{Language Group} & Room 40& 13/155& 0.618 & 0.008 & 0.690 & 0.156 & 0.681 & 0.051 \\
    & Room 41 & 2/53 & 0.886 & 0.703 & 1.019 & -0.563 & 0.799 & 0.427\\ 
    & Room 43 & 5/52 & 0.526 & -0.075 & 0.779 & -0.082 & 0.592 & -0.186 \\
    & Room 68 & 3/53 & 1.007 & -0.312 & 0.823 & -0.072 & 1.016 & -0.276 \\\hline
    \multirow{3}{*}{Math Group} & Room 40 & 7/80 & 0.671 & 0.247 & 0.849 & -0.327 & 0.640 & 0.178 \\
    & Room 41 & 9/110 & 0.763 & -0.114 & 0.715 & 0.044 & 0.783 & -0.185 \\
    & Room 43 & 7/101 & 0.657 & -0.046 & 0.753 & 0.197 & 0.768 & 0.030 \\\hline
    \multirow{8}{*}{Course} & Chapel & 11/12 & 0.938 & -0.268 & 1.100 & -0.297 & 0.693 & 0.021 \\
    & English & 18/71 & 0.599 & 0.255 & 0.484 & 0.057 & 0.639 & 0.340 \\
    & Health & 8/8 & 0.991 & 0.155 & 1.538 & 0.149 & 0.776 & -0.035 \\
    & Language & 20/38 & 0.986 & -0.206 & 0.970 & -0.512 & 0.975 & -0.372 \\
    & Maths & 20/79 & 0.551 & -0.033 & 0.816 & 0.150 & 0.779 & 0.0177 \\
    & PE & 8/8 & 1.044 & 0.235 & 1.314 & 0.103 & 0.845 & 0.224 \\
    & Politics & 19/43 & 0.754 & -0.119 & 0.784 & -0.101 & 0.643 & -0.341 \\
    & Science & 19/32 & 0.641 & 0.006 & 0.537 & 0.252 & 0.687 & -0.050\\\hline
    \end{tabular}}
    \label{tab:case_sty2_overall_stats}
\end{table}

We carried out a quantitative analysis to assess the potential negative impacts derived from neglecting certain contextual factors. The findings, detailed in Table~\ref{tab:case_study2_eval_results}, indicate that specific situated contexts -- such as thermal comfort, the group division (\eg language and math groups), and the courses students were engaged in prior to data collection -- significantly influence the performance of the prediction algorithm. For example, as illustrated in Table~\ref{tab:case_study2_eval_results}, the algorithm's ability to accurately assess emotional engagement was statistically different between students who were comfortable with the room temperature and those who were not (feeling either too cold or too warm). To delve deeper into this observation, we analyzed the mean squared error (MSE) of the regression algorithm across different levels of thermal comfort. As reported in Table~\ref{tab:case_sty2_overall_stats}, the algorithm showed a notably lower error rate ($MSE$ = 0.631, $p$ = 0.011) when predicting the emotional engagement of students who were comfortable with the temperature, compared to those who were not ($MSE$ = 0.822 for students feeling the temperature should be cooler and $MSE$ = 0.742 for students feeling the temperature should be warmer). Similarly, our analysis indicated a significantly higher error rate ($MSE$ = 0.849, $p$ = 0.043) in predicting the cognitive engagement of students in Room 40 for their math class, as opposed to those in other math groups ($MSE$ = 0.715 for Room 41 and $MSE$ = 0.753 for Room 43). Interestingly, our analysis revealed no evidence of algorithmic bias or harm, both with gender and in predicting student behavioral engagement.

\paragraph{\textbf{Additional Experiment on Bias Mitigation}} To determine if the findings from our first evaluation study can be replicated using the same bias mitigation technique -- incorporate context data into both the training and testing phase of the algorithm -- we conducted an additional experiment in this evaluation study. 

\begin{table}[t]
\renewcommand{\arraystretch}{1}
\caption{\textbf{Results After Implementing Bias Mitigation Techniques}. The outcomes following the inclusion of language group assignment data in the training and testing phases of the emotional engagement prediction algorithm.}
\begin{subtable}[t]{0.49\linewidth}
\centering
\resizebox{1\textwidth}{!}{\begin{tabular}{|l|l|rrr|}
\hline
\multicolumn{1}{|c|}{\multirow{2}{*}{\textbf{Contextual Factors}}} & \multicolumn{1}{c|}{\multirow{2}{*}{\textbf{Model Variables}}} & \multicolumn{3}{c|}{\cellcolor[HTML]{CCCCCC}\textbf{\makecell[c]{Emotional Engagement}}} \\\cline{3-5} 
\multicolumn{1}{|c|}{}& \multicolumn{1}{c|}{}  &  \makecell[c]{Coef.} & \makecell[c]{Std. Error} &  \makecell[c]{P>|z|}  \\
\hline
\multirow{3}{*}{Gender}& Intercept & -0.026  & 0.096 &  0.785 \\
& groups [T.Male] & 0.030 & 0.143 & 0.834 \\
& Group Var & 0.060 & 0.045 &  \\\hline
\multirow{4}{*}{Thermal Comfort}& Intercept &  -0.154 & 0.103 & 0.137 \\
& groups [T.No change] & 0.288 & 0.110 & \cellcolor[HTML]{F4CCCC}**\textbf{0.009} \\
& groups [T.Warmer] & -0.094 & 0.147 & 0.523 \\
& Group Var & 0.055 & 0.042 &  \\\hline
\multirow{5}{*}{\textbf{Language Group}}& Intercept  & 0.022 & 0.093 & 0.811 \\
& groups [T.Room 41] & 0.236 & 0.240 & 0.325 \\
& groups [T.Room 43] & -0.060 & 0.181 & 0.740 \\
& groups [T.Room 68] & -0.282 & 0.195 & 0.147 \\
& Group Var & 0.053 & 0.043 &  \\\hline
\end{tabular}}
\end{subtable}
\hspace{\fill}
\begin{subtable}[t]{0.49\linewidth}
\centering
\resizebox{1\textwidth}{!}{\begin{tabular}{|l|l|rrr|}
\hline
\multicolumn{1}{|c|}{\multirow{2}{*}{\textbf{Contextual Factors}}} & \multicolumn{1}{c|}{\multirow{2}{*}{\textbf{Model Variables}}} & \multicolumn{3}{c|}{\cellcolor[HTML]{CCCCCC}\textbf{\makecell[c]{Emotional Engagement}}} \\\cline{3-5} 
\multicolumn{1}{|c|}{}& \multicolumn{1}{c|}{}  &  \makecell[c]{Coef.} & \makecell[c]{Std. Error} &  \makecell[c]{P>|z|}  \\
\hline
\multirow{3}{*}{Math Room }& Intercept & 0.164 & 0.131 & 0.210 \\
& groups [T.Room 41] & -0.299 & 0.169 & 0.077 \\
& groups [T.Room 43] & -0.172 & 0.175 & 0.327 \\
& Group Var & 0.050 & 0.042 &   \\\hline
\multirow{9}{*}{Course}& Intercept & -0.345 & 0.228 &  0.131 \\
& groups [T.English] & 0.579 & 0.240 &  \cellcolor[HTML]{F4CCCC}*\textbf{0.016} \\
& groups [T.Health] & 0.554 & 0.351 &  0.115 \\
& groups [T.Language] & 0.111 & 0.254 & 0.663  \\
& groups [T.Maths] &  0.304 & 0.239 & 0.203 \\
& groups [T.PE] &  0.490 & 0.354 & 0.166  \\
& groups [T.Politics] & 0.202 & 0.251 & 0.420  \\
& groups [T.Science] & 0.320 & 0.261 & 0.219 \\
& Group Var &  0.051 & 0.040  &  \\\hline
\end{tabular}}
\end{subtable}
\label{tab:case_study2_bias_mitigation}
\end{table}

As an example, we aimed to mitigate the algorithmic bias caused by the lack of detailed information about students' assignments to different language groups, specifically in the context of predicting student emotional engagement. As indicated in Table~\ref{tab:case_study2_bias_mitigation}, incorporating this information into both the training and testing phases of the algorithm proved effective in reducing algorithmic harm. Compared to Table~\ref{tab:overall_evaluation}, this method resulted in a more equitable prediction performance across students assigned to various language groups. However, it was less effective in addressing biases related to different levels of thermal comfort and the variety of courses students were taking.

\section{Discussion}\label{sec:discussion}
In this section, we begin by summarizing the key insights we derived from our evaluation studies (Section \ref{sec:discussion_studies}). This summary covers the various findings, their implications, and how they contribute to our understanding of designing behavioral sensing technologies. Following this, we delve into a reflection on our framework, examining its strengths, limitations, and considering perspectives that extend beyond its current scope (Section \ref{sec:discussion_ethical}).

\subsection{Key Insights on Evaluation Studies}\label{sec:discussion_studies}

Our evaluations of two real-world behavioral well-being sensing technology studies demonstrated the practicality and effectiveness of our proposed framework. Throughout both evaluation studies, we identified a range of commonalities as well as unique findings, which we detail below.

\subsubsection{Potential Harms to Marginalized Groups Due to Context-insensitivity}
In both of our evaluation studies, we uncovered a critical and consistent issue with existing behavioral sensing technology designs: a widespread disregard for potential harms to users. Our evaluation, as detailed in Table~\ref{tab:overall_evaluation}, revealed that none of the designs thoroughly considered steps 2, 4, 5, and 6 proposed in our framework during their design processes. Furthermore, while a few designs did consider the collection of more diverse contextual datasets (\eg ~\cite{xu2019leveraging,xu2021leveraging}), this type of data was not utilized effectively during the algorithm training and testing phases. Our quantitative analysis of algorithm performance substantiates the concern of potential harms due to this oversight. Both studies identified significant issues, either identity-based harm or situation-based harm. Identity-based harm, which is more straightforward, can directly impact marginalized groups. In contrast, the concept of situation-based harm is more nuanced and can be subtle in its impact on these groups. To illustrate this further, in addition to the example discussed in Section~\ref{sec:situation-bias}, our second evaluation study provides another insightful instance. Specifically, we found that the algorithm for predicting emotional engagement was less effective for students who felt uncomfortably cold or warm compared to those who were comfortable with the temperature (as detailed in Tables~\ref{tab:case_study2_eval_results} and \ref{tab:case_sty2_overall_stats}). This finding may imply a potential indirect harm to individuals of lower socioeconomic status, who may have restricted access to air conditioning and thus are more likely to experience algorithmic harms~\cite{gronlund2014racial}.

\subsubsection{Balance Trade-offs in Achieving Algorithmic Fairness}\label{discussion:balance_stakeholder}

In our first evaluation study, as detailed in Section~\ref{sec:bias_mitigation}, we encountered a trade-off when attempting to mitigate harms. We observed that while using an in-processing mechanism (\ie incorporating the sensitive attribute into the training and testing process) helped reduce bias for that particular attribute, it unexpectedly introduced a bias towards other sensitive attributes. This outcome highlights the complex and nuanced nature of mitigating algorithmic harms. It suggests that while certain mitigation strategies might address specific biases, they can also unintentionally create new harms. 

Understanding these intricate trade-offs, technology builders should explicitly ask the question when developing context-sensitive algorithms: which group of users should be prioritized and to what extent? Our framework emphasized the importance of comprehensively understanding users' backgrounds and specific needs to answer the first part of this question. Our emphasis on engaging more with users and involving them throughout the design of the behavioral sensing technologies answers the latter part of this question.

The complex field of algorithmic fairness often presents trade-offs not only between different user groups but also among various fairness metrics~\cite{hardt2016equality}. Each metric provides a unique lens on bias, focusing on different aspects of equity. However, optimizing for one metric may lead to unintended negative outcomes in another, creating challenging scenarios~\cite{corbett2018measure}. For example, in our first evaluation study, when sensitive attributes were incorporated into Saeb \etal's algorithm training and testing, it reduced bias in accuracy disparity for most sensitive attributes, as shown in Table \ref{tab:tab_acc_w_demo_delta}. Yet, a detailed examination of other key fairness metrics like disparity in false negative rates and positive rates (Tables \ref{tab:tab_fnr_w_demo_delta} and \ref{tab:tab_fpr_w_demo_delta}) reveals significant variations. This highlights the complex dynamics involved in fairness optimization, where achieving fairness in one dimension might inadvertently lead to imbalances in others, emphasizing the multifaceted nature of achieving algorithmic fairness. This dynamic becomes even more crucial in behavioral sensing technologies, where data collection remains continuous, and system behavior is deeply adaptive to changing contexts. This recognition sets the stage for our subsequent discussion (Section \ref{sec:discussion_ethical}), delving into the essential requirement for regular and systematic monitoring. 

\subsubsection{A Need for Engage Users Throughout the Design Process}\label{sec:discussion_engagement}

Another key finding from both of our evaluation studies is the complete absence of user involvement throughout the design process of existing behavioral sensing technologies. Given the widespread use of behavioral sensing technologies, particularly in the mental health domain, this is concerning. As argued by Zhu \etal~\cite{zhu2018value}, engaging with users in the early stage of the design process can ensure that technologies are designed with a deep understanding of users' needs and values, which can significantly enhance user acceptance and satisfaction. Furthermore, the engagement of users extends beyond the initial design phase to include ongoing feedback loops. Regular interactions with users allow for iterative improvements and adjustments based on evolving needs, emerging challenges, and changing social contexts~\cite{amershi2019guidelines}. However, it is important to recognize the balance between involving users to mitigate technology harms and minimizing demands on their time and resources. This is especially necessary for people with different needs~\cite{zhang2022impact}. Striking this balance ensures that users' contributions are meaningful and sustainable, and that their valuable input genuinely shapes the direction of the technology while respecting their availability and capacity.

\subsection{Towards More Responsible Behavioral Sensing}\label{sec:discussion_ethical}

In this section, we discuss various aspects both within and beyond our current framework. These include the necessity of continuous maintenance for longitudinal behavioral sensing deployment while minimizing human labor, considerations of harms in other components of behavioral sensing technology, as well as the incorporation of other responsible considerations. Our intention is to inspire researchers and designers towards the conception and realization of more responsible behavioral sensing technologies.

\subsubsection{Need for Regular Maintenance while Alleviating Excessive Human Labor}
During our evaluation studies, we discovered that the reason behind the lack of continuous maintenance for responsible deployment (as outlined in step 6 of our framework) in these behavioral sensing technology designs was that the technologies were not truly deployed in real-world settings. This limitation arises from the nature of the limited datasets and the absence of deployable algorithmic systems~\cite{zhang2023framework,xu2023globem}. 

Behavioral sensing technologies in real-world applications operate in dynamic environments and depend on continuously evolving data streams. This dynamic nature increases the risk of situation-based harms, demanding continuous vigilance to guarantee that the system's accuracy and fair alignment persist over time. Regular maintenance is a key step in achieving this goal. By continuously updating deployment datasets, refining algorithms to accommodate temporal dynamics, and regularly monitoring the system's performance, the system can uphold its reliability and fulfill its ethical responsibility towards users and users. However, it is also important to avoid overburdening human resources with excessive maintenance demands. High human labor requirements can lead to operational inefficiencies, increased costs, and hindered scalability~\cite{frey2017future}. Striking a balance between rigorous maintenance and an approach that minimizes the burden on human resources is pivotal. Leveraging automation, intelligent monitoring, and adaptive algorithms can potentially alleviate this issue. 

\subsubsection{Other Aspects of Harms in Behavioral Sensing Technology}

In this work, our emphasis centers on addressing
harms within the algorithmic aspects of behavioral sensing technology. Nonetheless, it is important to acknowledge that considerations of harms should extend beyond the algorithm design process and include other critical components of the system. One dimension, for example, to focus on is the user interface and user interaction.

A user interface that is designed without incorporating the consideration of harms to users can inadvertently influence users to make certain choices or take specific actions. When these nudges disproportionately benefit particular groups, it can result in disparate outcomes that perpetuate inequality. Additionally, if user interfaces are not designed with accessibility in mind, individuals with disabilities might face barriers in accessing the interacting with the system. Future endeavors should take this aspect into account when designing their user interface. Finally, approaches to transparently informing users about potential fairness concerns, similar to transparent information about accuracy concerns, should be incorporated into a deployed fair behavioral sensing technology.

\subsubsection{Expanding Responsible Considerations to Address Additional Needs} 

While our study primarily concentrates on algorithmic harms in the context of behavioral sensing technology, it is essential to recognize that responsible considerations encompass a broader spectrum of dimensions, such as transparency, privacy, and accountability (\eg~\cite{liao2023ai,ehsan2021expanding}). As behavioral sensing technology becomes more widely used, ensuring transparency becomes a pressing concern. A lack of transparency can lead to opacity and a lack of user trust~\cite{buccinca2021trust,banovic2023being}. Interpretability and explanation techniques are crucial in addressing this issue, allowing users to understand algorithmic decisions and aiding system developers in identifying potential harms~\cite{doshi2017towards}. Furthermore, continuous data collection in behavioral sensing raises significant privacy challenges~\cite{reynolds2004affective}. The risk of unauthorized and unintended data sharing is ever-present. Researchers can develop privacy-preserving algorithms and techniques tailored specifically to behavioral sensing environments and delve into privacy-enhancing technologies, such as secure multi-party computation~\cite{goldreich1998secure,keller2020mp}, federated learning~\cite{rieke2020future}, and differential privacy~\cite{dwork2006differential}, and incorporate them into the framework. In addition, accountability is about establishing mechanisms to hold responsible parties accountable for the outcomes of their algorithms and systems~\cite{diakopoulos2016accountability}. In behavioral sensing, accountability can be challenging due to complex decision-making processes and interactions between various components. Technology builders of these technologies must be held answerable for their impact on users. Expanding our proposed framework to include all the above-discussed aspects can create a more comprehensive foundation for responsible behavioral sensing technology design and deployment.

\section{Limitations and Future Work}

While our research included two comprehensive evaluations of real-world behavioral sensing technologies across various domains and machine ML tasks, aiming to derive broader conclusions, we acknowledge that both evaluation studies are situated within the overarching theme of well-being prediction. This specific focus may limit the generalizability of our findings to other applications outside of well-being prediction. Furthermore, in both of our evaluation studies, we identified various instances of identity-based and situation-based harms. However, we note that some aspects might still be overlooked. Moreover, the datasets used in our evaluations presented their own set of constraints. For instance, specific instances of harms, such as disability status, were either unrepresented or underrepresented due to limited data collection or small sample sizes. This data limitation restricts our ability to make conclusive statements about these groups. Future work should continue to explore the potential harms in a broader array of behavioral sensing technology applications, identify additional instances both within the two potential harms we discussed and beyond, as well as collect more inclusive datasets for a more comprehensive analysis. Additionally, while our study demonstrates the feasibility of our proposed framework for evaluating and mitigating harms, its practical usability and broader application still require validation. Comparing our framework with others, although beyond the scope of this study, will be crucial in future meta-evaluations to benchmark its effectiveness and utility in the field.

\section{Conclusion}
\label{sec:conclusion}

In conclusion, this paper identified three gaps in the existing literature. We introduce a specific framework for evaluating and mitigating the context-induced harms in behavioral sensing. Our framework highlights a comprehensive consideration of potential broad and domain-specific harms due to a lack of context sensitivity and the need for iterative harm mitigation and continuous maintenance for responsible technology deployment. Through our two evaluation studies, we showcase the feasibility of our proposed framework. By conducting quantitative analyses, we uncover empirical evidence of identity-based and situation-based harms in existing behavioral sensing technologies and validate the framework's capability in identifying and mitigating these harms. We discuss the insights learned from the evaluation studies and other aspects within and beyond the scope of our proposed framework. We hope our work inspires technology builders in our field to amplify their attention to the significance of incorporating harm considerations and other responsible considerations in behavioral sensing technologies.

\bibliographystyle{ACM-Reference-Format}
\bibliography{main}

\newpage

\appendix
\section{Appendix}\label{appendix}

\subsection{Percentage of Each Group within Each Sensitive Attribute}

\begin{figure}[htb!]
    \centering
    \caption{Percentage of each group within each sensitive attribute. The protected group for each sensitive attribute (\eg first-gen) is shaded in dark colors, while the unprotected group is shaded in light colors (\eg non-first-gen). Non-male includes women, transgender individuals, and genderqueer individuals, non-heterosexual includes homosexual, bisexual, and asexual individuals, and non-white includes black, asian, latinx, and biracial.}
    \includegraphics[scale=0.23]{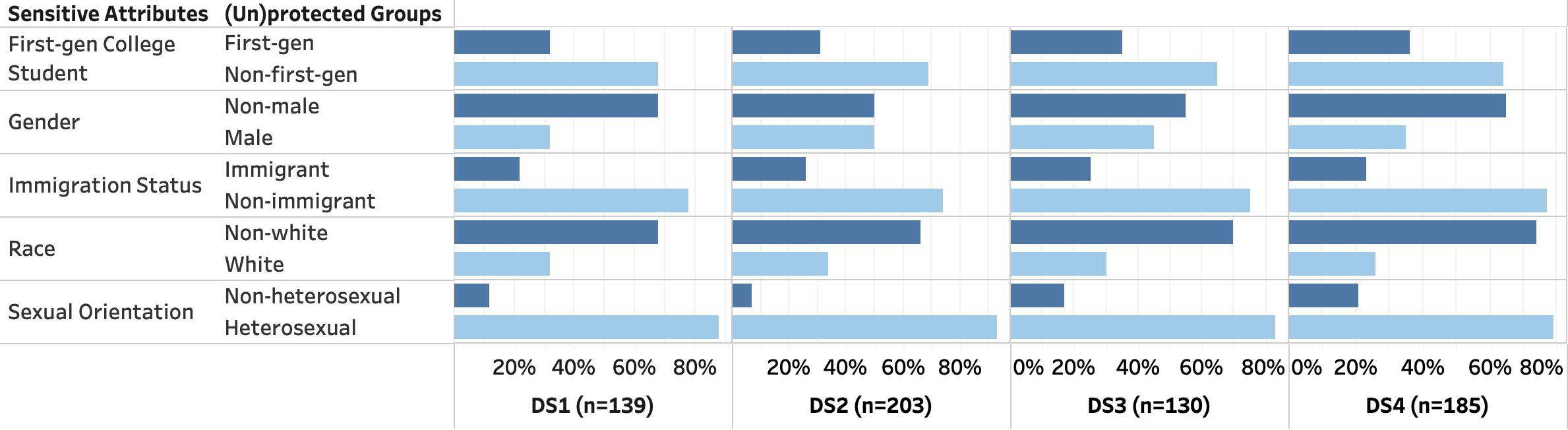}
    \label{fig:demo}
\end{figure}



\subsection{Case Study 1: Example of the Statistical Evaluation and Experimental Implementation}\label{appdx:example}
\subsubsection{Example of the Statistical Evaluation}

\begin{figure}[htb!]
    \centering
    \caption{Example of fairness evaluation based on the disparities in accuracy, false negative rate, and false positive rate. (a) shows the synthetic data for 20 individuals, with 6 belonging to the protected group (represented by ``x'' marks) and 14 belonging to the unprotected group (represented by ``$\boldsymbol{\cdot}$'' marks). (b) visualizes the distribution and disparities of predictions for both groups, where correction predictions are depicted in green and incorrect predictions in red.}
     \subfloat[Synthetic data for 20 individuals.]{\includegraphics[scale=0.14]{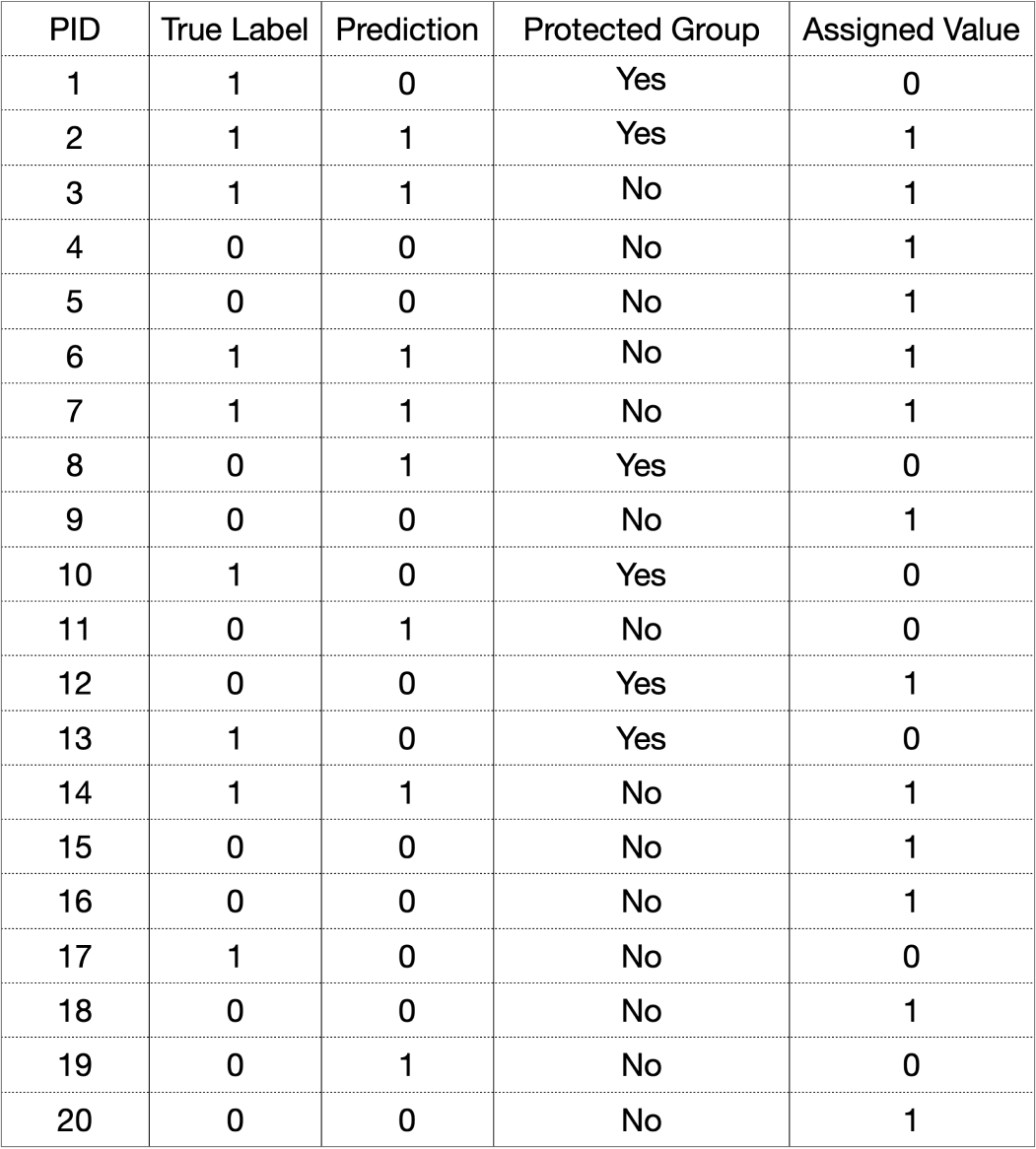}
    \label{fig:example_data}}
    \subfloat[Visualization of prediction distributions and disparities.]{ \includegraphics[scale=0.30]{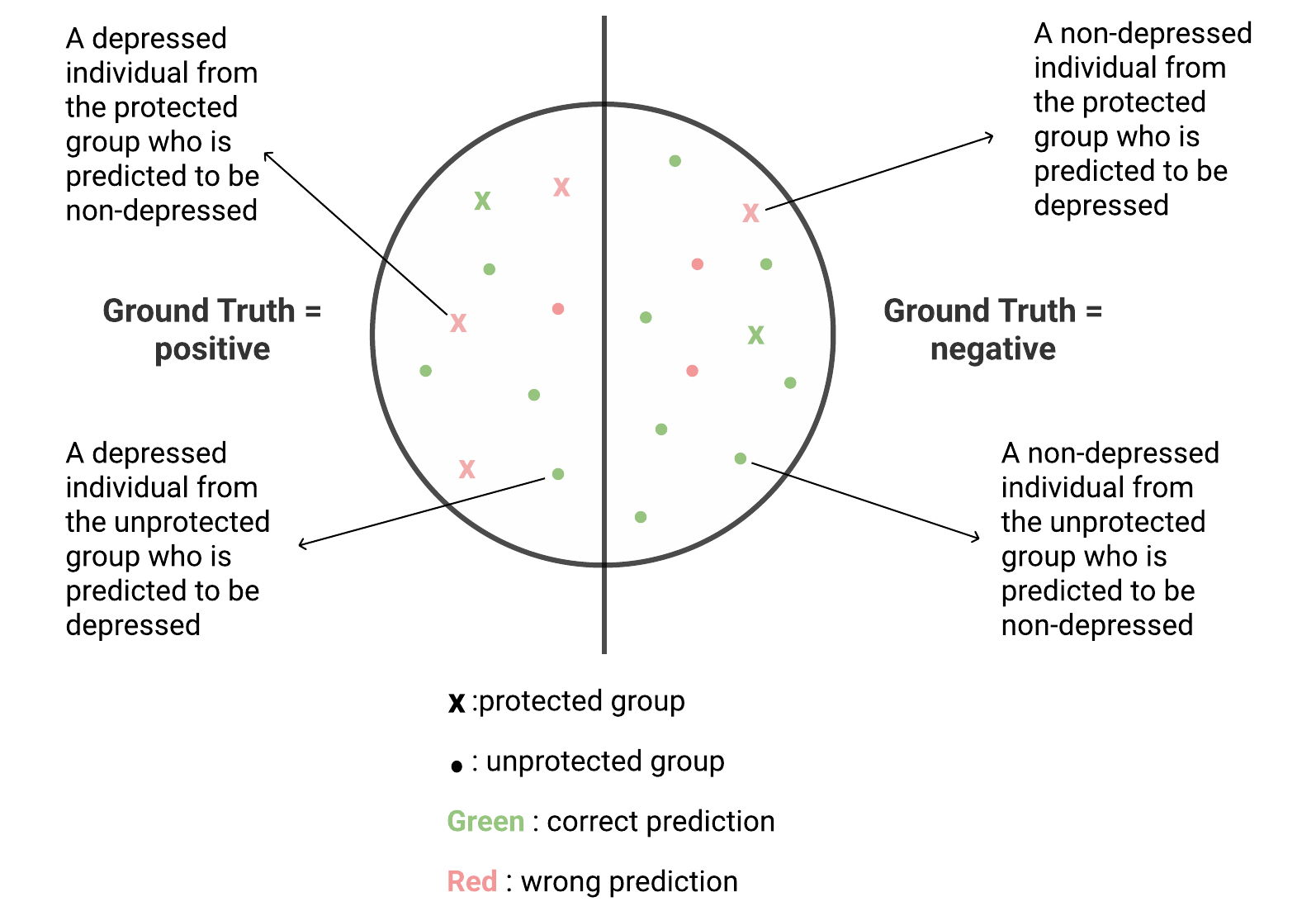}
    \label{fig:disparity}}
    \label{fig:example_disparity}  
\end{figure}

In Figure \ref{fig:example_disparity}, we present an illustrative example to demonstrate our approach to fairness evaluation. In this example, we generated synthesized ground-truth labels and predictions from an algorithm for a sample of 20 individuals. Among these individuals, 6 are part of the protected group, while 14 belong to the unprotected group (as shown in Figure \ref{fig:example_data}). We assigned a value of ``1'' for accurate predictions and ``0'' for inaccurate predictions based on the correctness of the predictions. 

Figure \ref{fig:disparity} visualizes the distribution of predictions for both the protected group (represented by ``x'') and the unprotected group (represented by ``$\boldsymbol{\cdot}$''). The circle in the figure represents the distribution of predictions, with the left side indicating cases where all ground truth values are positive (representing individuals with depression in our case study), and the right side representing cases where all ground truth values are negative (representing individuals without depression in our case study). The accuracy of the predictions is indicated by the color, with green representing correct predictions by the algorithm and red representing incorrect predictions. 

In this example, when considering the disparity in accuracy, the algorithm made incorrect predictions for 4 out of 6 individuals from the protected group (represented by the red ``x'' marks among all both red and green ``x'' marks). Conversely, for the unprotected group, the algorithm made incorrect predictions for 3 out of 14 individuals (depicted by the red ``$\boldsymbol{\cdot}$'' marks among all red and green ``$\boldsymbol{\cdot}$'' marks). To assess the statistical significance of these disparities, we conducted the Mann-Whitney U test in combination with the Benjamini-Hochberg correction. Specifically, we applied this test to the 2 ``1'' values and 4 ``0'' values corresponding to the protected group, as well as the 11 ``1'' values and 3 ``0'' values corresponding to the unprotected group.

When examining the difference in false negative rates, the relevant information for statistical analysis is contained in the left portion of the circle depicted in Figure \ref{fig:disparity}. Specifically, we conducted a statistical test on the 1 ``1'' value and 3 ``0'' values in the protected group, as well as the 4 ``1'' values and 1 ``0'' value in the unprotected group. Similarly, an evaluation of the disparity in false positive rates was conducted on the marks on the right side of the circle in Figure \ref{fig:disparity}.

\subsubsection{Experimental Implementation}
We applied the two evaluation criteria as defined in Section~\ref{sec:define_eval} to evaluate the fairness of the eight depression detection algorithms. We provide a detailed explanation of our statistical analyses to capture disparities in accuracy, false negative rate, and false positive rate below (an example of this approach can be found in above). We provide open access to our evaluation codebase to enable reference and reproducibility for future research.

To perform the Benjamini-Hochberg correction, we first calculated the $p$ values for all attributes using the Mann-Whitney U test. Then, we arranged the $p$ values in ascending order and assign ranks to them, with the smallest $p$ value receiving rank 1, the second smallest receiving rank 2, and so on. Next, we calculated the adjusted $q$ values for each individual $p$ value using the formula: $(i/m) \times Q$, where $i$ is the rank of the individual $p$ value, $m$ is the total number of tests, and $Q$ is the false discovery rate, 0.05. Finally, we compared the original $p$ values to the calculated $q$ values. Attributes with a $p$ value smaller than the corresponding $q$ value and less than 0.05 were considered to have significant differences (which we highlighted in red in Tables \ref{tab_all_w_t_demo} and \ref{tab:xu_interpretable_first-gen}).

To examine potential disparities across various groups of one algorithm, we employed a systematic approach. Initially, we categorized algorithm predictions based on their correctness, assigning a value of ``1'' to instances where the algorithm accurately predicted the ground truth and a value of ``0'' to instances where the algorithm falsely predicted the ground truth. Subsequently, we applied the Mann-Whitney test in conjunction with the Benjamini-Hochberg correction to different subsets of the ``0'' and ``1'' values to evaluate the following three hypotheses. First, we conducted a thorough analysis to determine whether the algorithms exhibited comparable accuracy in predicting the ground truth for both the protected and unprotected groups, aiming to evaluate potential disparities in accuracy. To achieve this, we performed the Mann-Whitney test with the Benjamini-Hochberg correction on the complete set of ``0'' and ``1'' values. Second, our assessment focused on whether the algorithms demonstrated similar false negative rates in predicting the ground truth for both the protected and unprotected groups, to identify potential disparities in false negative rates. To accomplish this, we conducted the same test on the subset of ``0'' and ``1'' values where the ground truth labels were positive. Similarly, we proceeded to investigate whether the algorithms displayed comparable false positive rates for both the protected and unprotected groups. This was achieved by applying the same test on the subset of ``0'' and ``1'' values where the ground truth labels were negative. 

\clearpage

\subsection{Comparisons of Depression Scores for Different Groups of Four Datasets.}

\begin{figure}[htb!]
    \centering 
    \caption{Comparisons of depression (BDI-II) scores for different groups of four datasets. The red dotted line indicates the cutoff point (i.e., 13) for BDI-II scores, which is used to distinguish between students with at least mild depressive symptoms (BDI-II score >=13) and those without (BDI-II < 13). Significance levels after Benjamini-Hochberg (B-H) correction are marked with an asterisk (*$p<0.05$) in red on the subplot. First-gen, BA, and HET represent first-generation college students, bachelor, and heterosexual, respectively.}\label{fig_BDI2}
    \subfloat[DS1 (Year 2018, Pre-COVID)]{\includegraphics[scale=0.35]{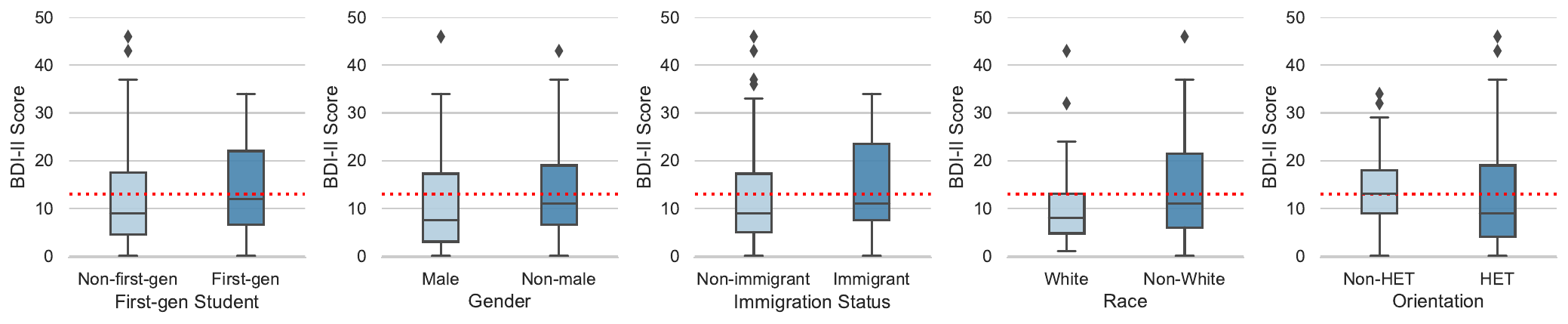}
    \label{fig:boxplot1}}
    
    \subfloat[DS2 (Year 2019, Pre-COVID)]{\includegraphics[scale=0.35]{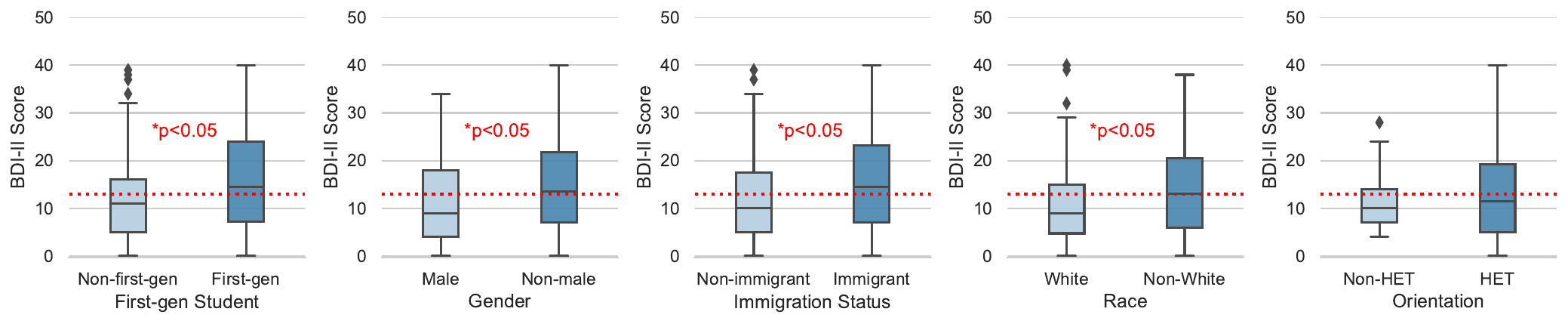}
    \label{fig:boxplot2}}

    \subfloat[DS3 (Year 2020, COVID Year)]{\includegraphics[scale=0.35]{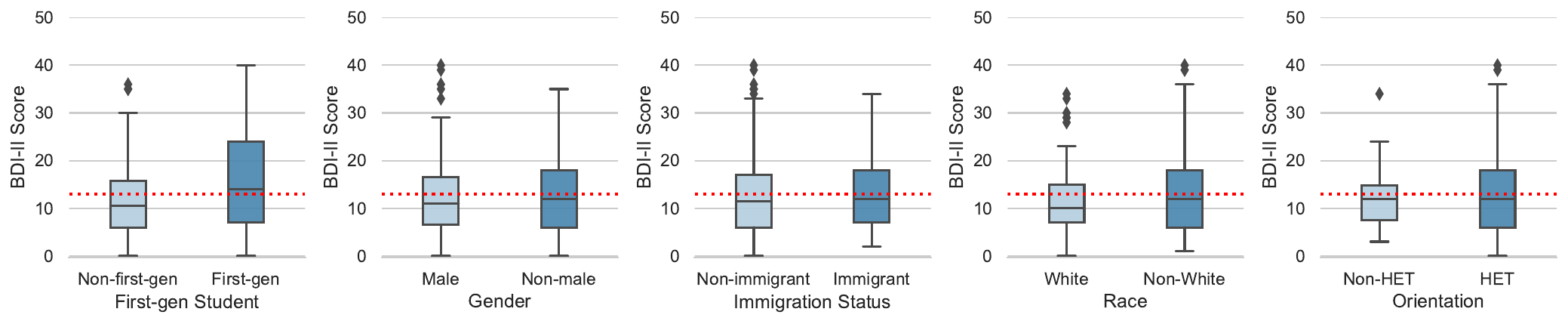}
    \label{fig:boxplot3}}
    
    \subfloat[DS4 (Year 2021, COVID Year)]{\includegraphics[scale=0.35]{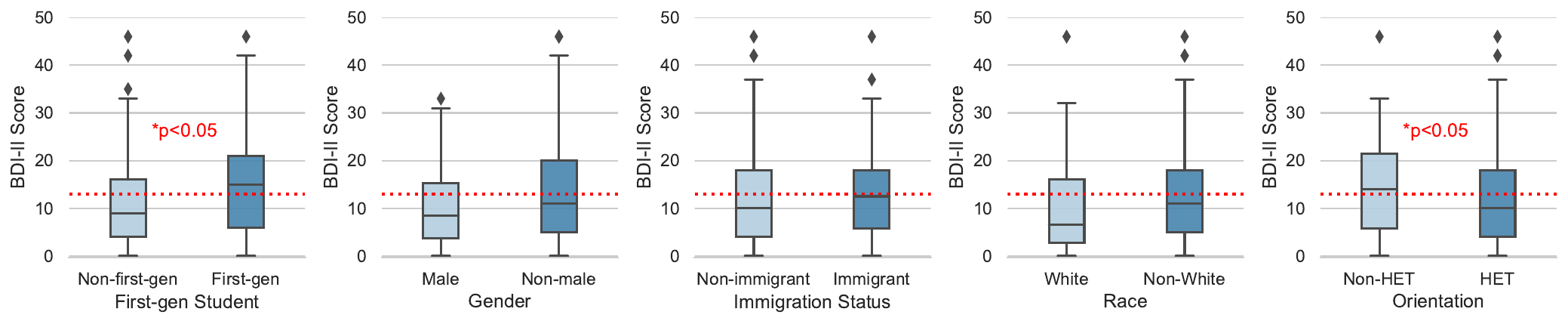}
    \label{fig:boxplot4}}
\end{figure}

\clearpage


\begin{table}[htb!]
\caption{Summary of bias changes with the addition of sensitive attributes in the training and testing process in terms of \textbf{disparity in accuracy}. This table provides an overview of bias alterations resulting from the inclusion of sensitive attributes during the training and testing processes, using disparity in accuracy as the fairness metrics. It encompasses bias amplification and reduction for each sensitive attribute across the four datasets. The comparison highlights the consequences of adding or excluding sensitive attributes in training and testing. Extra bias is denoted in red, while reduced bias is highlighted in green. For instance, considering the Xu\_interpretable algorithm, Tables \ref{tab_all_w_t_demo} and \ref{tab:xu_interpretable_first-gen} present fairness evaluation outcomes before and after incorporating data related to first-generation college student status. When this sensitive attribute is introduced, an additional bias towards gender emerges in DS4, indicated by the label ``1'' in this table.}
\renewcommand{\arraystretch}{1}
\resizebox{0.8\textwidth}{!}
{\begin{tabular}{|c|l|ccccc|}
\hline
\multicolumn{1}{|c|}{}& \multicolumn{1}{c|}{}& \multicolumn{5}{c|}{\textbf{Sensitive Attributes}}\\ \cline{3-7}
\multicolumn{1}{|c|}{\multirow{-2}{*}{\textbf{Algorithm}}} & \multicolumn{1}{c|}{\multirow{-2}{*}{\textbf{Added Attributes}}} & \multicolumn{1}{c}{\makecell[c]{First-gen \\ college student}} & \multicolumn{1}{c}{Gender} & \multicolumn{1}{c}{\makecell[c]{Immigration \\ Status}} & \multicolumn{1}{c}{Race} & \multicolumn{1}{c|}{\makecell[c]{Sexual \\Orientation}} \\ \hline
\multicolumn{1}{|c|}{}& First-gen college student & \cellcolor[HTML]{F4CCCC}1 & \cellcolor[HTML]{D9EAD3}-1& 0 & 0 & 0 \\
\multicolumn{1}{|c|}{}& Gender & \cellcolor[HTML]{F4CCCC}1 & \cellcolor[HTML]{F4CCCC}1 & 0 & 0 & 0 \\
\multicolumn{1}{|c|}{}& Immigration Status & \cellcolor[HTML]{F4CCCC}1 & \cellcolor[HTML]{D9EAD3}-1& 0 & 0 & 0 \\
\multicolumn{1}{|c|}{}& Race & 0& \cellcolor[HTML]{D9EAD3}-1& 0 & 0 & 0 \\
\multicolumn{1}{|c|}{\multirow{-5}{*}{Wahle \etal}}& Sexual Orientation & \cellcolor[HTML]{F4CCCC}1 & 0& 0 & 0 & 0 \\ \hline
 & First-gen college student & \cellcolor[HTML]{D9EAD3}-1& \cellcolor[HTML]{D9EAD3}-1& \cellcolor[HTML]{D9EAD3}-1 & \cellcolor[HTML]{D9EAD3}-1 & \cellcolor[HTML]{D9EAD3}-1 \\
 & Gender & \cellcolor[HTML]{D9EAD3}-2& \cellcolor[HTML]{F4CCCC}1 & \cellcolor[HTML]{D9EAD3}-1 & \cellcolor[HTML]{D9EAD3}-1 & \cellcolor[HTML]{D9EAD3}-1 \\
 & Immigration Status & 0& 0& \cellcolor[HTML]{D9EAD3}-1 & \cellcolor[HTML]{D9EAD3}-1 & 0 \\
 & Race & 0& 0& 0 & 0 & 0 \\
\multirow{-5}{*}{Saeb \etal}& Sexual Orientation & \cellcolor[HTML]{D9EAD3}-1& 0& 0 & \cellcolor[HTML]{D9EAD3}-1 & \cellcolor[HTML]{D9EAD3}-1 \\\hline
 & First-gen college student & 0& 0& 0 & 0 & 0 \\
 & Gender & 0& \cellcolor[HTML]{F4CCCC}1 & 0 & 0 & 0 \\
 & Immigration Status & 0& 0& \cellcolor[HTML]{F4CCCC}1& 0 & \cellcolor[HTML]{F4CCCC}1\\
 & Race & 0& \cellcolor[HTML]{D9EAD3}-1& 0 & \cellcolor[HTML]{F4CCCC}1& \cellcolor[HTML]{F4CCCC}1\\
\multirow{-5}{*}{Farhan \etal} & Sexual Orientation & 0& \cellcolor[HTML]{F4CCCC}1 & 0 & 0 & 0 \\\hline
 & First-gen college student & \cellcolor[HTML]{F4CCCC}1 & 0& 0 & 0 & \cellcolor[HTML]{D9EAD3}-1 \\
 & Gender & 0& 0& 0 & 0 & 0 \\
 & Immigration Status & \cellcolor[HTML]{D9EAD3}-1& \cellcolor[HTML]{D9EAD3}-1& 0 & 0 & 0 \\
 & Race & 0& 0& 0 & 0 & 0 \\
\multirow{-5}{*}{Canzian \etal}& Sexual Orientation & 0& \cellcolor[HTML]{D9EAD3}-1& \cellcolor[HTML]{F4CCCC}2& 0 & 0 \\\hline
 & First-gen college student & \cellcolor[HTML]{F4CCCC}1 & 0& \cellcolor[HTML]{D9EAD3}-1 & \cellcolor[HTML]{F4CCCC}1& \cellcolor[HTML]{D9EAD3}-1 \\
 & Gender & 0& \cellcolor[HTML]{F4CCCC}1 & 0 & \cellcolor[HTML]{F4CCCC}1& 0 \\
 & Immigration Status & 0& 0& \cellcolor[HTML]{D9EAD3}-1 & \cellcolor[HTML]{F4CCCC}1& 0 \\
 & Race & 0& \cellcolor[HTML]{F4CCCC}1 & \cellcolor[HTML]{D9EAD3}-1 & 0 & 0 \\
\multirow{-5}{*}{Wang \etal}& Sexual Orientation & \cellcolor[HTML]{F4CCCC}1 & \cellcolor[HTML]{F4CCCC}1 & \cellcolor[HTML]{D9EAD3}-1 & 0 & 0 \\\hline
 & First-gen college student & 0& \cellcolor[HTML]{F4CCCC}1 & 0 & \cellcolor[HTML]{F4CCCC}2& \cellcolor[HTML]{F4CCCC}1\\
 & Gender & \cellcolor[HTML]{D9EAD3}-1& 0& 0 & \cellcolor[HTML]{F4CCCC}1& 0 \\
 & Immigration Status & 0& 0& 0 & \cellcolor[HTML]{F4CCCC}1& 0 \\
 & Race & 0& 0& 0 & \cellcolor[HTML]{F4CCCC}1& 0 \\
\multirow{-5}{*}{Lu \etal}& Sexual Orientation & \cellcolor[HTML]{D9EAD3}-1& 0& 0 & 0 & \cellcolor[HTML]{F4CCCC}1\\\hline
 & First-gen college student & 0& \cellcolor[HTML]{F4CCCC}1 & 0 & 0 & 0 \\
 & Gender & 0& 0& 0 & 0 & 0 \\
 & Immigration Status & 0& \cellcolor[HTML]{F4CCCC}1 & 0 & 0 & 0 \\
 & Race & 0& \cellcolor[HTML]{F4CCCC}1 & 0 & 0 & 0 \\
\multirow{-5}{*}{Xu\_interpretable \etal} & Sexual Orientation & 0& \cellcolor[HTML]{F4CCCC}1 & 0 & 0 & 0 \\\hline
 & First-gen college student & 0& 0& 0 & 0 & 0 \\
 & Gender & 0& 0& 0 & 0 & 0 \\
 & Immigration Status & 0& 0& 0 & 0 & 0 \\
 & Race & 0& 0& 0 & 0 & 0 \\
\multirow{-5}{*}{Xu\_personalized \etal}& Sexual Orientation & 0& 0& 0 & 0 & 0 \\\hline
\end{tabular}}\label{tab:tab_acc_w_demo_delta}
\end{table}

\begin{table}[htb!]
\caption{Summary of bias changes with the addition of sensitive attributes in the training and testing process in terms of \textbf{disparity in false negative rates}. This table provides an overview of bias alterations resulting from the inclusion of sensitive attributes during the training and testing processes, using disparity in false negative rates as the fairness metrics. It encompasses bias amplification and reduction for each sensitive attribute across the four datasets. The comparison highlights the consequences of adding or excluding sensitive attributes in training and testing. Extra bias is denoted in red, while reduced bias is highlighted in green.}
\renewcommand{\arraystretch}{1}
\resizebox{0.8\textwidth}{!}
{\begin{tabular}{|c|l|ccccc|}
\hline
\multicolumn{1}{|c|}{\cellcolor[HTML]{FFFFFF}} & \multicolumn{1}{c|}{\cellcolor[HTML]{FFFFFF}}   & \multicolumn{5}{c|}{\cellcolor[HTML]{FFFFFF}\textbf{Sensitive Attribute}} \\ \cline{3-7} 
\multicolumn{1}{|c|}{\multirow{-2}{*}{\cellcolor[HTML]{FFFFFF}\textbf{Algorithm}}} & \multicolumn{1}{c|}{\multirow{-2}{*}{\cellcolor[HTML]{FFFFFF}\textbf{Added Attribute}}} & \multicolumn{1}{c}{\cellcolor[HTML]{FFFFFF}\makecell[c]{First-gen \\ college student}} & \multicolumn{1}{c}{\cellcolor[HTML]{FFFFFF}Gender} & \multicolumn{1}{c}{\cellcolor[HTML]{FFFFFF}\makecell[c]{Immigration \\ Status}} & \multicolumn{1}{c}{\cellcolor[HTML]{FFFFFF}Race} & \multicolumn{1}{c|}{\cellcolor[HTML]{FFFFFF}\makecell[c]{Sexual \\ Orientation}} \\ \hline
\multicolumn{1}{|c|}{\cellcolor[HTML]{FFFFFF}} & First-gen college student     & \cellcolor[HTML]{F4CCCC}2     & 0  & 0    & \cellcolor[HTML]{F4CCCC}1    & 0    \\
\multicolumn{1}{|c|}{\cellcolor[HTML]{FFFFFF}} & Gender    & 0   & 0  & 0    & 0& 0    \\
\multicolumn{1}{|c|}{\cellcolor[HTML]{FFFFFF}} & Immigration Status    & 0   & 0  & \cellcolor[HTML]{F4CCCC}1  & \cellcolor[HTML]{F4CCCC}1    & 0    \\
\multicolumn{1}{|c|}{\cellcolor[HTML]{FFFFFF}} & Race      & 0   & 0  & 0    & 0& 0    \\
\multicolumn{1}{|c|}{\multirow{-5}{*}{\cellcolor[HTML]{FFFFFF}Wahle \etal}}       & Sexual Orientation    & 0   & 0  & 0    & 0& \cellcolor[HTML]{F4CCCC}1  \\ \hline
\cellcolor[HTML]{FFFFFF} & First-gen college student     & \cellcolor[HTML]{F4CCCC}3     & \cellcolor[HTML]{F4CCCC}2      & 0    & \cellcolor[HTML]{F4CCCC}2    & \cellcolor[HTML]{D9EAD3}-2 \\
\cellcolor[HTML]{FFFFFF} & Gender    & \cellcolor[HTML]{F4CCCC}1     & \cellcolor[HTML]{F4CCCC}2      & 0    & \cellcolor[HTML]{F4CCCC}1    & \cellcolor[HTML]{D9EAD3}-1 \\
\cellcolor[HTML]{FFFFFF} & Immigration Status    & 0   & \cellcolor[HTML]{F4CCCC}1      & 0    & 0& \cellcolor[HTML]{D9EAD3}-1 \\
\cellcolor[HTML]{FFFFFF} & Race      & 0   & 0  & 0    & \cellcolor[HTML]{F4CCCC}2    & \cellcolor[HTML]{D9EAD3}-1 \\
\multirow{-5}{*}{\cellcolor[HTML]{FFFFFF}Saeb \etal}  & Sexual Orientation    & 0   & 0  & 0    & 0& 0    \\\hline
\cellcolor[HTML]{FFFFFF} & First-gen college student     & \cellcolor[HTML]{F4CCCC}1     & 0  & \cellcolor[HTML]{D9EAD3}-1 & 0& \cellcolor[HTML]{F4CCCC}1  \\
\cellcolor[HTML]{FFFFFF} & Gender    & 0   & 0  & \cellcolor[HTML]{D9EAD3}-1 & 0& 0    \\
\cellcolor[HTML]{FFFFFF} & Immigration Status    & 0   & \cellcolor[HTML]{F4CCCC}1      & \cellcolor[HTML]{F4CCCC}2  & 0& \cellcolor[HTML]{F4CCCC}1  \\
\cellcolor[HTML]{FFFFFF} & Race      & 0   & 0  & \cellcolor[HTML]{D9EAD3}-1 & 0& 0    \\
\multirow{-5}{*}{\cellcolor[HTML]{FFFFFF}Farhan \etal}& Sexual Orientation    & 0   & 0  & 0    & 0& \cellcolor[HTML]{F4CCCC}1  \\\hline
\cellcolor[HTML]{FFFFFF} & First-gen college student     & \cellcolor[HTML]{F4CCCC}2     & \cellcolor[HTML]{F4CCCC}1      & 0    & 0& 0    \\
\cellcolor[HTML]{FFFFFF} & Gender    & 0   & 0  & 0    & 0& 0    \\
\cellcolor[HTML]{FFFFFF} & Immigration Status    & 0   & 0  & \cellcolor[HTML]{F4CCCC}3  & \cellcolor[HTML]{F4CCCC}1    & \cellcolor[HTML]{F4CCCC}1  \\
\cellcolor[HTML]{FFFFFF} & Race      & \cellcolor[HTML]{F4CCCC}1     & 0  & 0    & \cellcolor[HTML]{F4CCCC}1    & 0    \\
\multirow{-5}{*}{\cellcolor[HTML]{FFFFFF}Canzian \etal}       & Sexual Orientation    & 0   & 0  & 0    & 0& \cellcolor[HTML]{F4CCCC}1  \\\hline
\cellcolor[HTML]{FFFFFF} & First-gen college student     & \cellcolor[HTML]{F4CCCC}1     & 0  & 0    & \cellcolor[HTML]{F4CCCC}1    & 0    \\
\cellcolor[HTML]{FFFFFF} & Gender    & 0   & 0  & 0    & 0& 0    \\
\cellcolor[HTML]{FFFFFF} & Immigration Status    & 0   & 0  & 0    & 0& 0    \\
\cellcolor[HTML]{FFFFFF} & Race      & 0   & 0  & 0    & \cellcolor[HTML]{F4CCCC}1    & 0    \\
\multirow{-5}{*}{\cellcolor[HTML]{FFFFFF}Wang \etal}  & Sexual Orientation    & 0   & 0  & 0    & 0& 0    \\\hline
\cellcolor[HTML]{FFFFFF} & First-gen college student     & \cellcolor[HTML]{F4CCCC}1     & \cellcolor[HTML]{F4CCCC}1      & 0    & \cellcolor[HTML]{F4CCCC}1    & 0    \\
\cellcolor[HTML]{FFFFFF} & Gender    & 0   & \cellcolor[HTML]{F4CCCC}2      & 0    & 0& 0    \\
\cellcolor[HTML]{FFFFFF} & Immigration Status    & \cellcolor[HTML]{D9EAD3}-1    & \cellcolor[HTML]{F4CCCC}1      & \cellcolor[HTML]{F4CCCC}1  & 0& 0    \\
\cellcolor[HTML]{FFFFFF} & Race      & \cellcolor[HTML]{D9EAD3}-1    & 0  & 0    & \cellcolor[HTML]{F4CCCC}3    & 0    \\
\multirow{-5}{*}{\cellcolor[HTML]{FFFFFF}Lu \etal}    & Sexual Orientation    & \cellcolor[HTML]{D9EAD3}-1    & 0  & 0    & 0& 0    \\\hline
\cellcolor[HTML]{FFFFFF} & First-gen college student     & \cellcolor[HTML]{D9EAD3}-1    & 0  & 0    & 0& 0    \\
\cellcolor[HTML]{FFFFFF} & Gender    & 0   & \cellcolor[HTML]{F4CCCC}1      & 0    & 0& 0    \\
\cellcolor[HTML]{FFFFFF} & Immigration Status    & \cellcolor[HTML]{D9EAD3}-1    & 0  & 0    & 0& 0    \\
\cellcolor[HTML]{FFFFFF} & Race      & \cellcolor[HTML]{D9EAD3}-1    & 0  & 0    & 0& 0    \\
\multirow{-5}{*}{\cellcolor[HTML]{FFFFFF}Xu\_interpretable \etal} & Sexual Orientation    & \cellcolor[HTML]{D9EAD3}-1    & 0  & 0    & 0& 0    \\\hline
\cellcolor[HTML]{FFFFFF} & First-gen college student     & 0   & 0  & 0    & 0& 0    \\
\cellcolor[HTML]{FFFFFF} & Gender    & 0   & 0  & 0    & 0& 0    \\
\cellcolor[HTML]{FFFFFF} & Immigration Status    & 0   & 0  & 0    & 0& 0    \\
\cellcolor[HTML]{FFFFFF} & Race      & 0   & 0  & 0    & \cellcolor[HTML]{D9EAD3}0    & 0    \\
\multirow{-5}{*}{\cellcolor[HTML]{FFFFFF}Xu\_personalized \etal}  & Sexual Orientation    & 0   & 0  & 0    & \cellcolor[HTML]{D9EAD3}0    & 0 \\\hline  
\end{tabular}}\label{tab:tab_fnr_w_demo_delta}
\end{table}

\begin{table}[htb!]
\caption{Summary of bias changes with the addition of sensitive attributes in the training and testing process in terms of \textbf{disparity in false positive rates}. This table provides an overview of bias alterations resulting from the inclusion of sensitive attributes during the training and testing processes, using disparity in false positive rates as the fairness metrics. It encompasses bias amplification and reduction for each sensitive attribute across the four datasets. The comparison highlights the consequences of adding or excluding sensitive attributes in training and testing. Extra bias is denoted in red, while reduced bias is highlighted in green.}
\renewcommand{\arraystretch}{1}
\resizebox{0.8\textwidth}{!}
{\begin{tabular}{|c|l|ccccc|}
\hline
\rowcolor[HTML]{FFFFFF} 
\multicolumn{1}{|c|}{\cellcolor[HTML]{FFFFFF}}     & \multicolumn{1}{c|}{\cellcolor[HTML]{FFFFFF}}  & \multicolumn{5}{c|}{\cellcolor[HTML]{FFFFFF}\textbf{Sensitive Attribute}}\\ \cline{3-7} 
\rowcolor[HTML]{FFFFFF} 
\multicolumn{1}{|c|}{\multirow{-2}{*}{\cellcolor[HTML]{FFFFFF}\textbf{Algorithm}}} & \multicolumn{1}{c|}{\multirow{-2}{*}{\cellcolor[HTML]{FFFFFF}\textbf{Added Attribute}}} & \multicolumn{1}{c}{\cellcolor[HTML]{FFFFFF}\makecell[c]{First-gen \\college student}} & \multicolumn{1}{c}{\cellcolor[HTML]{FFFFFF}Gender} & \multicolumn{1}{c}{\cellcolor[HTML]{FFFFFF}\makecell[c]{Immigration \\ Status}} & \multicolumn{1}{c}{\cellcolor[HTML]{FFFFFF}Race} & \multicolumn{1}{c|}{\cellcolor[HTML]{FFFFFF}\makecell[c]{Sexual \\ Orientation}} \\ \hline
\rowcolor[HTML]{FFFFFF} 
\multicolumn{1}{|c|}{\cellcolor[HTML]{FFFFFF}}     & First-gen college student     & \cellcolor[HTML]{F4CCCC}3    & 0  & \cellcolor[HTML]{F4CCCC}1      & \cellcolor[HTML]{F4CCCC}1        & 0    \\
\rowcolor[HTML]{FFFFFF} 
\multicolumn{1}{|c|}{\cellcolor[HTML]{FFFFFF}}     & Gender        & \cellcolor[HTML]{D9EAD3}-1   & \cellcolor[HTML]{F4CCCC}1 & 0    & 0& 0    \\
\rowcolor[HTML]{FFFFFF} 
\multicolumn{1}{|c|}{\cellcolor[HTML]{FFFFFF}}     & Immigration Status   & 0  & 0  & \cellcolor[HTML]{F4CCCC}1      & 0& 0    \\
\rowcolor[HTML]{FFFFFF} 
\multicolumn{1}{|c|}{\cellcolor[HTML]{FFFFFF}}     & Race & \cellcolor[HTML]{F4CCCC}1    & 0  & 0    & \cellcolor[HTML]{F4CCCC}2        & 0    \\
\rowcolor[HTML]{FFFFFF} 
\multicolumn{1}{|c|}{\multirow{-5}{*}{\cellcolor[HTML]{FFFFFF}Wahle \etal}}       & Sexual Orientation   & 0  & 0  & 0    & \cellcolor[HTML]{F4CCCC}1        & \cellcolor[HTML]{F4CCCC}1      \\ \hline
\rowcolor[HTML]{FFFFFF} 
\cellcolor[HTML]{FFFFFF} & First-gen college student     & \cellcolor[HTML]{F4CCCC}4    & \cellcolor[HTML]{D9EAD3}-1& 0    & \cellcolor[HTML]{F4CCCC}1        & 0    \\
\rowcolor[HTML]{F4CCCC} 
\cellcolor[HTML]{FFFFFF} & \cellcolor[HTML]{FFFFFF}Gender& 1  & 2  & 1    & 2& 1    \\
\rowcolor[HTML]{F4CCCC} 
\cellcolor[HTML]{FFFFFF} & \cellcolor[HTML]{FFFFFF}Immigration Status     & 1  & 1  & 2    & 3& 1    \\
\rowcolor[HTML]{FFFFFF} 
\cellcolor[HTML]{FFFFFF} & Race & \cellcolor[HTML]{F4CCCC}1    & 0  & \cellcolor[HTML]{F4CCCC}1      & \cellcolor[HTML]{F4CCCC}1        & 0    \\
\rowcolor[HTML]{FFFFFF} 
\multirow{-5}{*}{\cellcolor[HTML]{FFFFFF}Saeb \etal}     & Sexual Orientation   & 0  & \cellcolor[HTML]{D9EAD3}-1& 0    & 0& \cellcolor[HTML]{F4CCCC}1      \\\hline
& First-gen college student     & \cellcolor[HTML]{F4CCCC}1    & \cellcolor[HTML]{F4CCCC}1 & \cellcolor[HTML]{FFFFFF}0      & \cellcolor[HTML]{F4CCCC}1        & \cellcolor[HTML]{FFFFFF}0      \\
& Gender        & \cellcolor[HTML]{FFFFFF}0    & \cellcolor[HTML]{F4CCCC}2 & \cellcolor[HTML]{FFFFFF}0      & \cellcolor[HTML]{F4CCCC}1        & \cellcolor[HTML]{D9EAD3}-1     \\
& Immigration Status   & \cellcolor[HTML]{FFFFFF}0    & \cellcolor[HTML]{FFFFFF}0 & \cellcolor[HTML]{F4CCCC}1      & \cellcolor[HTML]{D9EAD3}-1       & \cellcolor[HTML]{D9EAD3}-1     \\
& Race & \cellcolor[HTML]{FFFFFF}0    & \cellcolor[HTML]{FFFFFF}0 & \cellcolor[HTML]{FFFFFF}0      & \cellcolor[HTML]{D9EAD3}-1       & \cellcolor[HTML]{D9EAD3}-1     \\
\multirow{-5}{*}{Farhan \etal}    & Sexual Orientation   & \cellcolor[HTML]{FFFFFF}0    & \cellcolor[HTML]{F4CCCC}1 & \cellcolor[HTML]{FFFFFF}0      & \cellcolor[HTML]{D9EAD3}-1       & \cellcolor[HTML]{D9EAD3}-1     \\\hline
& First-gen college student     & \cellcolor[HTML]{F4CCCC}1    & \cellcolor[HTML]{D9EAD3}-1& \cellcolor[HTML]{F4CCCC}1      & \cellcolor[HTML]{FFFFFF}0        & \cellcolor[HTML]{D9EAD3}-1     \\
& Gender        & \cellcolor[HTML]{FFFFFF}0    & \cellcolor[HTML]{FFFFFF}0 & \cellcolor[HTML]{FFFFFF}0      & \cellcolor[HTML]{FFFFFF}0        & \cellcolor[HTML]{FFFFFF}0      \\
& Immigration Status   & \cellcolor[HTML]{D9EAD3}-1   & \cellcolor[HTML]{D9EAD3}-1& \cellcolor[HTML]{F4CCCC}3      & \cellcolor[HTML]{F4CCCC}1        & \cellcolor[HTML]{D9EAD3}-1     \\
& Race & \cellcolor[HTML]{FFFFFF}0    & \cellcolor[HTML]{FFFFFF}0 & \cellcolor[HTML]{FFFFFF}0      & \cellcolor[HTML]{D9EAD3}-1       & \cellcolor[HTML]{FFFFFF}0      \\
\multirow{-5}{*}{Canzian \etal}   & Sexual Orientation   & \cellcolor[HTML]{D9EAD3}-1   & \cellcolor[HTML]{D9EAD3}-1& \cellcolor[HTML]{F4CCCC}1      & \cellcolor[HTML]{FFFFFF}0        & \cellcolor[HTML]{FFFFFF}0      \\\hline
& First-gen college student     & \cellcolor[HTML]{F4CCCC}2    & \cellcolor[HTML]{FFFFFF}0 & \cellcolor[HTML]{FFFFFF}0      & \cellcolor[HTML]{FFFFFF}0        & \cellcolor[HTML]{D9EAD3}-1     \\
& Gender        & \cellcolor[HTML]{FFFFFF}0    & \cellcolor[HTML]{FFFFFF}0 & \cellcolor[HTML]{FFFFFF}0      & \cellcolor[HTML]{FFFFFF}0        & \cellcolor[HTML]{D9EAD3}-1     \\
& Immigration Status   & \cellcolor[HTML]{FFFFFF}0    & \cellcolor[HTML]{FFFFFF}0 & \cellcolor[HTML]{F4CCCC}1      & \cellcolor[HTML]{FFFFFF}0        & \cellcolor[HTML]{FFFFFF}0      \\
& Race & \cellcolor[HTML]{F4CCCC}1    & \cellcolor[HTML]{F4CCCC}1 & \cellcolor[HTML]{FFFFFF}0      & \cellcolor[HTML]{F4CCCC}1        & \cellcolor[HTML]{D9EAD3}-1     \\
\multirow{-5}{*}{Wang \etal}      & Sexual Orientation   & \cellcolor[HTML]{FFFFFF}0    & \cellcolor[HTML]{FFFFFF}0 & \cellcolor[HTML]{FFFFFF}0      & \cellcolor[HTML]{FFFFFF}0        & \cellcolor[HTML]{D9EAD3}-1     \\\hline
\rowcolor[HTML]{F4CCCC} 
\cellcolor[HTML]{FFFFFF} & \cellcolor[HTML]{FFFFFF}First-gen college student       & 3  & 1  & \cellcolor[HTML]{D9EAD3}-1     & 1& 1    \\
\rowcolor[HTML]{FFFFFF} 
\cellcolor[HTML]{FFFFFF} & Gender        & 0  & \cellcolor[HTML]{F4CCCC}3 & \cellcolor[HTML]{D9EAD3}-1     & 0& \cellcolor[HTML]{D9EAD3}-1     \\
\cellcolor[HTML]{FFFFFF} & \cellcolor[HTML]{FFFFFF}Immigration Status     & \cellcolor[HTML]{F4CCCC}1    & \cellcolor[HTML]{FFFFFF}0 & \cellcolor[HTML]{F4CCCC}1      & \cellcolor[HTML]{F4CCCC}1        & \cellcolor[HTML]{D9EAD3}-1     \\
\rowcolor[HTML]{FFFFFF} 
\cellcolor[HTML]{FFFFFF} & Race & \cellcolor[HTML]{F4CCCC}2    & 0  & 0    & \cellcolor[HTML]{F4CCCC}2        & \cellcolor[HTML]{D9EAD3}-1     \\
\rowcolor[HTML]{FFFFFF} 
\multirow{-5}{*}{\cellcolor[HTML]{FFFFFF}Lu \etal}& Sexual Orientation   & 0  & 0  & 0    & 0& 0    \\\hline
\rowcolor[HTML]{FFFFFF} 
\cellcolor[HTML]{FFFFFF} & First-gen college student     & 0  & \cellcolor[HTML]{F4CCCC}1 & 0    & 0& 0    \\
\rowcolor[HTML]{FFFFFF} 
\cellcolor[HTML]{FFFFFF} & Gender        & 0  & 0  & 0    & 0& 0    \\
\rowcolor[HTML]{FFFFFF} 
\cellcolor[HTML]{FFFFFF} & Immigration Status   & \cellcolor[HTML]{F4CCCC}1    & \cellcolor[HTML]{F4CCCC}1 & 0    & 0&      \\
\rowcolor[HTML]{FFFFFF} 
\cellcolor[HTML]{FFFFFF} & Race & \cellcolor[HTML]{F4CCCC}1    & \cellcolor[HTML]{F4CCCC}1 & 0    & 0& 0    \\
\rowcolor[HTML]{FFFFFF} 
\multirow{-5}{*}{\cellcolor[HTML]{FFFFFF}Xu\_interpretable \etal} & Sexual Orientation   & \cellcolor[HTML]{F4CCCC}1    & \cellcolor[HTML]{F4CCCC}1 & 0    & 0& 0    \\ \hline
\rowcolor[HTML]{FFFFFF} 
\cellcolor[HTML]{FFFFFF} & First-gen college student     & 0  & 0  & 0    & \cellcolor[HTML]{D9EAD3}-1       & 0    \\
\rowcolor[HTML]{FFFFFF} 
\cellcolor[HTML]{FFFFFF} & Gender        & 0  & 0  & 0    & \cellcolor[HTML]{D9EAD3}-1       & 0    \\
\rowcolor[HTML]{FFFFFF} 
\cellcolor[HTML]{FFFFFF} & Immigration Status   & 0  & 0  & 0    & \cellcolor[HTML]{D9EAD3}-1       & 0    \\
\rowcolor[HTML]{FFFFFF} 
\cellcolor[HTML]{FFFFFF} & Race & 0  & 0  & 0    & \cellcolor[HTML]{D9EAD3}-1       & 0    \\
\rowcolor[HTML]{FFFFFF} 
\multirow{-5}{*}{\cellcolor[HTML]{FFFFFF}Xu\_personalized \etal}  & Sexual Orientation   & 0  & 0  & 0    & \cellcolor[HTML]{D9EAD3}-1       & 0   \\\hline
\end{tabular}}\label{tab:tab_fpr_w_demo_delta}
\end{table}

\end{document}